# Harnessing Ambient Sensing & Naturalistic Driving Systems to Understand Links Between Driving Volatility and Crash Propensity in School Zones – A generalized hierarchical mixed logit framework

Behram Wali, Ph.D.
Lead Research Scientist,
Urban Design 4 Health, Inc.
bwali@ud4h.com
Postdoctoral Scholar,
Sensenable City Lab,
Massachusetts Institute of Technology,
77 Massachusetts Avenue,
Cambridge, MA 02139, United States.

Asad J. Khattak, Ph.D.
Beaman Distinguished Professor
Transportation Program Coordinator
Department of Civil & Environmental Engineering
University of Tennessee,
Knoxville, TN 37996, USA
akhattak@utk.edu

# Harnessing Ambient Sensing & Naturalistic Driving Systems to Understand Links Between Driving Volatility and Crash Propensity in School Zones – A generalized hierarchical mixed logit framework

**Abstract –** With the advent of seemingly unstructured big data, and through seamless integration of computation and physical components, cyber-physical systems (CPS) provide an innovative way to enhance safety and resiliency of transport infrastructure. This study focuses on real-world microscopic driving behavior and its relevance to school zone safety – expanding the capability, usability, and safety of dynamic physical systems through data analytics. Driving behavior and school zone safety is a public health concern. The sequence of instantaneous driving decisions and its variations prior to involvement in safety critical events, defined as driving volatility, can be a leading indicator of safety. By harnessing unique naturalistic data on more than 41,000 normal, crash, and near-crash events featuring over 9.4 million temporal samples of real-world driving, a characterization of volatility in microscopic driving decisions is sought at school and non-school zone locations. A big data analytic methodology is proposed for quantifying driving volatility in microscopic real-world driving decisions. Eight different volatility measures are then linked with detailed event-specific characteristics, health history, driving history/experience, and other factors to examine crash propensity at school zones. A comprehensive yet fully flexible state-of-the-art generalized mixed logit framework is employed to fully account for distinct yet related methodological issues of scale and random heterogeneity, containing multinomial logit, random parameter logit, scaled logit, hierarchical scaled logit, and hierarchical generalized mixed logit as special cases. The results reveal that both for school and non-school locations, drivers exhibited greater intentional volatility prior to safety-critical events. Evidence is found that an increase in volatility in positive and negative vehicular jerk in longitudinal and lateral direction increases the probability of unsafe outcomes (crashes or near-crashes) at school zones. A one-unit increase in intentional volatility associated with positive vehicular jerk in longitudinal direction increases the probability of crash outcome by 0.0528 units. Importantly, the effect of negative vehicular jerk (braking) in longitudinal direction on the likelihood of crash outcome is almost double. Methodologically, Hierarchical Generalized Mixed Logit model resulted in best-fit, simultaneously accounting for scale and random heterogeneity. When accounted for separately, more parsimonious models accounting for scale heterogeneity performed comparably to the less parsimonious counterparts accounting for random heterogeneity. Importantly, even after accounting for random heterogeneity, substantial heterogeneity due to a "pure scale-effect" is still observed, underscoring the importance of scale effects in influencing the overall contours of variations in modeled relationships. The study demonstrates the value of observational study design and big data analytics for understanding extreme driving behaviors in safe vs. unsafe driving outcomes at vulnerable locations. Implications for designing personalized school zone behavioral countermeasures are discussed.

## 1. INTRODUCTION & BACKGROUND

Emerging technologies such as sensor-based monitoring, telematics, video and radar surveillance have enabled the monitoring of dynamic physical systems, generating countless terabytes of microscopic data about transport system performance (Katrakazas et al. 2015, Khattak and Wali 2017, Shrestha et al. 2017, Shladover 2018, Ganin et al. 2019). The advent of these technologies, along with the generation of seemingly unstructured big data, has established the elemental foundation of cyber-physical systems (CPS), allowing enhancement of transport system resiliency and safety in new and unique ways (Wu et al. 2014,

Shladover 2018). When integrated with computational advances and novel data analytic techniques, major societal challenges such as road safety can be addressed (Shi and Abdel-Aty 2015, Abdel-Aty et al. 2016, Imprialou and Quddus 2017, Li et al. 2017, Wali et al. 2018b, He et al. 2019, Tselentis et al. 2019). While tremendous progress has been made in advancing CPS technologies, the demand for transportation innovation across critical application domains (such as road safety) is driving the exigency to accelerate context-specific fundamental transportation research. As complex layers of urban networks and digital information blanket the urban landscape, new innovative techniques to the study of major transportation challenges are needed. By harnessing the big data generated by CPS technologies, this study focuses on real-world microscopic driving behavior and its relevance to school zone safety – expanding the capability, usability, and safety of dynamic physical systems through data analytics. In particular, a unique multidimensional naturalistic database is assembled for the analysis of microscopic driving behavior in normal as well as safety-critical (crashes/near-crashes) events at vulnerable locations such as school zones.

Recent statistics suggest that more than 90 percent of traffic crashes are influenced in a major way by driver behavior (Wang et al. 2013, FHWA 2017). The role of driving behavior can be more pronounced at vulnerable locations, such as school zones, where safety-critical interactions between motorized and non-motorized users are more likely. School zones have high concentration of pedestrian/bicyclist activity, and a relatively high proportion of children as well (Warsh et al. 2009, Ellison et al. 2013). Among all motor vehicle-pedestrian collisions, the area density of collisions is highest in school zones and decrease as distance from schools increased (Warsh et al. 2009). Importantly, within school zones, collisions were found to be more likely to occur among five to nine-year-old children (Warsh et al. 2009, Ellison et al. 2013). Driving behavior, as part of human error component, of passing motorists in school zones can be a critical component of overall pedestrian bicyclists' safety outcomes in school zones. In the U.S., approximately 33% of drivers displayed unsafe behaviors in school zones, whereas 1 in 10 drivers are found to be distracted in school zones (https://www.safekids.org/). This has resulted in a growing concern over the safety of school-aged children in addition to other vulnerable non-motorized road users in school zones.

For several decades, researchers have attempted to understand the behavioral correlates of crash risk or crash propensity at school zones. Typically, the focus is to examine the effect of presence of school zone on driver behavior measures, accident frequency (Strawderman et al. 2015), and subsequently speed compliance at school zones (Kattan et al. 2011). Likewise, by considering areas immediately surrounding schools, the relationships between physical, social attributes, pedestrian-vehicular crash risk (and/or the injuries sustained) have been analyzed (Clifton and Kreamer-Fults 2007). Overall, previous studies have generated useful knowledge critical to school zone safety countermeasure development (Clifton and Kreamer-Fults 2007, Kattan et al. 2011, Strawderman et al. 2015). However, important conceptual gaps remain. First, previous studies do not shed light on the real-world/naturalistic microscopic driving tasks and/or driver decisions that typically precede drivers' involvement in unsafe events. Second, actual driving behavior in school zones and its correlation with crash propensity is rarely examined. Third, previous analyses are primarily based on questionnaire surveys, controlled experiments, and/or police-reported crash data. Thus, it is important to generate new knowledge regarding the sequence of microscopic instantaneous driving decisions (e.g., speed, acceleration/deceleration, vehicular jerk, etc.) preceding driver's involvement in an unsafe outcome at school zones. An analysis of such a nature was not possible until very recent mainly due to data unavailability.

Thanks to rapid CPS technological advancements in recent years, countless terabytes of real-world data about vehicle and human movement is now a reality (Hankey et al. 2016, Chen et al. 2017, Liu and Khattak 2018, Zhang and Khattak 2018, Arbabzadeh et al. 2019, Arvin et al. 2019a, Ghasemzadeh and Ahmed 2019, Khattak et al. 2019, Yang et al. 2019). The main research issue is to use real world driving data to extract useful driving behavior information to enhance safety in school zones. Relevant in this regard is the concept of "driving volatility" that captures the extent of variations in driving, especially hard accelerations/braking and jerky maneuvers, and frequent switching between different safety-critical driving regimes (Kamrani et al. 2017, Khattak and Wali 2017, Wali et al. 2018b), and the references therein. The fundamental premise is that through monitoring and analysis of real-world instrumented data generated by CPS technologies, proactive approach to road safety can be formulated by giving warnings and alerts to drivers and which can reduce such volatility potentially improving safety.

## 1.1. Research Gap, Objectives & Contribution

Owing to the above prevalent gaps in the literature, the research questions in the present study are: 1) What pre-crash behaviors lead to risky outcomes in school zones where exposure is high, 2) What is the magnitude of driving volatility (both longitudinal and lateral) in school zones and non-school zones, and 3) How to appropriately quantify the correlations between driving volatility and crash propensity (involvement in crash and near-crash events) in school zones. In characterizing crash propensity, both crashes and near-crashes are considered vis-à-vis baseline/normal driving events. In particular, from a broader unsafe outcome perspective, consideration of near-crashes in quantifying crash propensity at a particular location (such as school zones) is important, as such "close calls" may foreshadow actual future crashes. To achieve the study objectives, the study harnesses a rigorous observational study design to help compare real-world microscopic driving decisions in normal vs. unsafe outcomes at school zones. In particular, the study builds upon the Second Strategic Highway Research Program's unique and largest Naturalistic Driving Study database of thousands of real-world driving events, in which a driver was involved in a normal, near-crash, or crash event. For over 40,000 naturalistic driving events, large-scale microscopic driving data pertaining to normal and unsafe outcomes are analyzed, and a rigorous data analytic methodology is developed to quantify volatility in microscopic driving decisions prior to safe/un-safe events, thus termed "event-based volatility". Careful attention is given to the issue of intentional vs. unintentional volatility (discussed later in detail). Once generated, the volatility indices are then linked with a broad spectrum of event-specific characteristics, health history, driving history/experience related factors, pre-event maneuvers/behaviors, secondary tasks, and roadway factors.

Once the unique multidimensional database is assembled, magnitudes of event-based driving volatility in longitudinal and lateral direction in school and non-school zones are examined. Then, advanced statistical models are developed to relate crash propensity at school zones with event-based volatility and several other observed/unobserved factors, generating new knowledge critical to the formulation of proactive warnings and alerts in case an unsafe outcome is anticipated in school zones. In this regard, we believe that methodological issues related to unobserved heterogeneity and omitted variable bias should be properly accounted for in analyses of such a nature. That is, it is important to control for unobserved factors that may influence unsafe outcomes at school zones but are not observed in data. If such unobserved factors could be included in a model, the correlations between driving volatility and unsafe outcomes can change, e.g., the magnitude or statistical significance of the relationship can change. While a broad spectrum of studies in the transport literature has successfully focused on capturing unobserved heterogeneity, much of the attention however has been on conceptualizing the heterogeneous associations in the form of random heterogeneity. It is important to trace the origin of unobserved heterogeneity, whether random or scale effects (or a combination of both) are likely influencing the contours of unobserved heterogeneity, and to disentangle the two related yet distinct issues (more details later)[1]. The key idea is that the presence of unobserved factors in the data could lead to heterogenous associations between 'observed factors' and crash propensity in school zones. Random and scale heterogeneity methods capture these potential heterogeneous associations in different ways (detailed discussion later). Random heterogeneity treatments allow estimation of a vector of (safety event-specific) β parameter estimates on specific exogenous factors (e.g., volatility measures) by assuming a certain distribution in the population. Whereas, scale heterogeneity methods (as implemented in this study) capture the heterogeneous associations by a pure scale effect (i.e., across safety events, all β estimates are scaled up or down in tandem) – implying that mechanisms leading to unsafe events could simply be more random in some cases than others (i.e., holding β estimates fixed, the scale of their error term is greater). From a methodological perspective, the present study contributes by developing state-of-the-art discrete outcome models based on generalized mixed logit framework (a superset of multinomial logit, random parameter logit, scaled logit, hierarchical scaled logit, and a generalized hierarchical scaled logit with random parameters) to link driving volatility with school zone crash propensity, accounting for scale and random heterogeneity in a single framework, with notable extension to account for the observed and unobserved components of the earlier. To the best of our knowledge, this is the first reported application of such a flexible discrete outcome framework in the context under discussion.

---

[1] In the context of the impacts of exogenous variables on driver injury severity using traditional General Estimates System (GES) database, a recent study carefully investigated whether the potential heterogeneous associations between exogenous factors and injury severity could be better represented through a scaled or random heterogeneity treatment in an ordered discrete framework (Marcoux et al. 2018). In doing so, a scaled ordered logit model was compared with a mixed (random parameter) ordered logit model concluding the statistical superiority of the earlier in terms of data fit – i.e., much of the heterogeneity in the associations can be captured by a pure scale effect. However, as the study acknowledged, mixed generalized ordered logit and scaled generalized ordered logit models were estimated separately precluding a simultaneous examination of scale and random heterogeneity (Marcoux et al. 2018).

## 2. METHODOLOGY

### 2.1. Data

Detailed microscopic data on driving decisions are needed to quantify and understand driving volatility prior to involvement in safety critical events (Kamrani et al. 2017, Khattak and Wali 2017, Wali et al. 2018b). The recently concluded SHRP2 Naturalistic Driving Study provides relevant data (Hankey et al. 2016). In this largest naturalistic driving study performed to date, the driving behaviors of approximately 3,400 participant drivers were recorded with over 4,300 years of naturalistic driving data collected between 2010 and 2013[2] (Hankey et al. 2016). The study data was collected from six naturalistic driving sites around the United States, with largest data collection sites in Florida, New York, North Carolina, Washington, Indiana, and Pennsylvania (Hankey et al. 2016). The study used approximately 3,300 participant vehicles (Hankey et al. 2016).

Out of the many data categories collected in the SHRP 2 NDS project, the data used in this study are extracted from "event data", "continuous motion data", "driving history questionnaire", and "medical conditions and medications". A total of 41,479 driving events (1,877 crashes, 6,881 near-crashes, and 32,721 baselines) are analyzed in this study. In particular, the framework consists of a populated table of safety critical events and baseline events, ranging from 20 seconds long to 30 seconds long. For baseline/normal driving and safety critical events (crashes/near-crashes), 20 and 30 seconds of microscopic driving data (speed, acceleration) are generally available, respectively. However, note that in many cases the seconds of data available for baselines and safety-critical events is less than 20 and 30 seconds respectively (discussed later). The sensor-based vehicle kinematics data are sampled at a 10 frames/second. Figure 1 summarizes the key large-scale seemingly disparate information used in this study, including elements of "event data", "driving history questionnaire" and "medical conditions and medications" (Figure 1).

---

[2] Over 4300 years refer to the total minutes of driving data collected as part of the SHRP Naturalistic Driving Study. Per the official documentation and InSight website (Hankey et al. 2016), a total of 6,559,367 trip files were collected for approximately 3,400 participant drivers. A trip file usually encompassed a whole trip from approximately 30 seconds after the ignition was turned on until the ignition was turned off. Adding the durations (minutes) of each of the 6,559,367 trips together would lead to over 4,300 years of driving data. However, the duration data are not available to the authors.

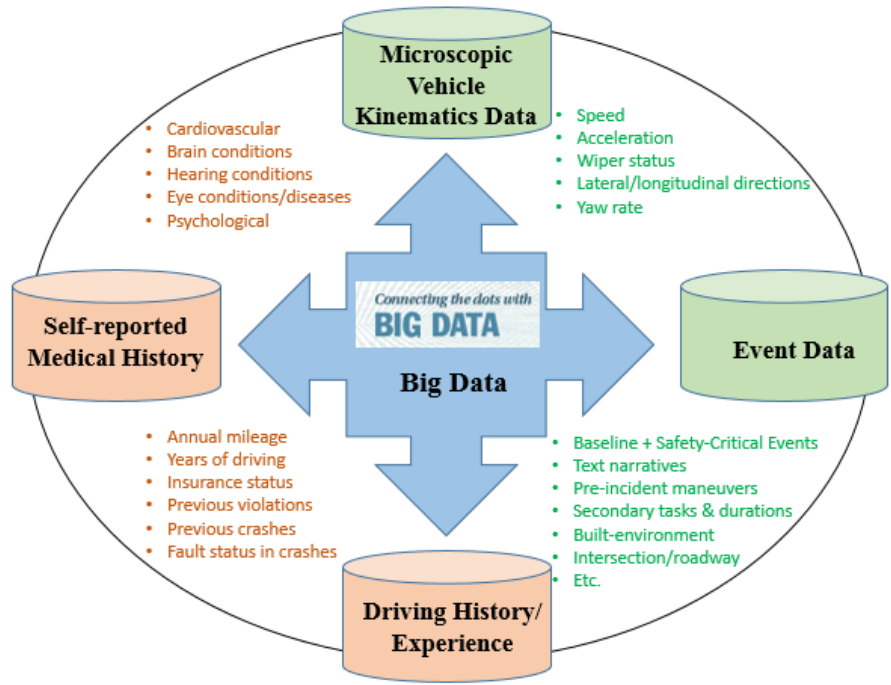

**FIGURE 1: Integration of Traditional and Emerging Transportation Data sources.**

The breakdown of data (41,520 events) across different states is provided in Table 1 – with the sampled events geographically distributed across the six states. Around 23.5% and 24.2% of the total events are sampled from Florida and Washington, respectively. Whereas, the least share of events is observed from Pennsylvania (around 6.6% of the total events) (Table 1).

TABLE 1: Distribution of Sampled Events Across U.S. States

| Location/State | Event Type | Number of Events | Percentage of Total Events | State-specific Percentage of Total Events |
|---|---|---|---|---|
| Florida | Additional Baseline | 2922 | 7.04 | 23.50 |
| Florida | Crash | 561 | 1.35 | |
| Florida | Near-Crash | 1640 | 3.95 | |
| Florida | Balanced-Sample Baseline | 4633 | 11.16 | |
| New York | Additional Baseline | 2696 | 6.49 | 22.22 |
| New York | Crash | 402 | 0.97 | |
| New York | Near-Crash | 1561 | 3.76 | |
| New York | Balanced-Sample Baseline | 4567 | 11.00 | |
| North Carolina | Additional Baseline | 1975 | 4.76 | 15.93 |
| North Carolina | Crash | 307 | 0.74 | |
| North Carolina | Near-Crash | 978 | 2.36 | |
| North Carolina | Balanced-Sample Baseline | 3355 | 8.08 | |
| Washington | Additional Baseline | 2910 | 7.01 | 24.29 |
| Washington | Crash | 430 | 1.04 | |
| Washington | Near-Crash | 2194 | 5.28 | |
| Washington | Balanced-Sample Baseline | 4551 | 10.96 | |
| Indiana | Additional Baseline | 943 | 2.27 | 7.41 |
| Indiana | Crash | 161 | 0.39 | |
| Indiana | Near-Crash | 392 | 0.94 | |
| Indiana | Balanced-Sample Baseline | 1580 | 3.81 | |
| Pennsylvania | Additional Baseline | 1137 | 2.74 | 6.65 |
| Pennsylvania | Crash | 92 | 0.22 | |
| Pennsylvania | Near-Crash | 221 | 0.53 | |
| Pennsylvania | Balanced-Sample Baseline | 1312 | 3.16 | |
| *All Six States* | *Total* | *41520* | *100* | *100.00* |

**Notes:** (1) The breakdown of sampled events across U.S. states was generously provided by Ms. Whitney Atkins (Virginia Tech Transportation Institute) upon request of the authors; (2) Note that data on 41,479 driving events (out of the total 41,520 events) are available for the present study.

### 2.2. Components and Calculation of Driving Volatility

The high-resolution vehicle kinematics data collected each one-tenth of a second are not useful to drivers and/or safety analysts in its raw form. While sensor-based driving data have recently become ubiquitously available, thanks to SHRP 2 and connected vehicles test beds across the U.S., techniques to extract valuable safety-critical information however from such data are not well-established. As such, by using big data analytic techniques, a unique aspect of the current study is to develop a methodology by which we can make sense of important but unstructured driving data. The end goal is to combine traditional and emerging data sources in a meaningful way critical to development of proactive safety tools for school zones.

As discussed earlier, driving volatility captures the extent of variations in driving, especially hard accelerations/braking and jerky maneuvers, and frequent switching between safety-critical driving regimes (Kamrani et al. 2017, Khattak and Wali 2017, Wali et al. 2018b). Driving volatility indices can shed light on microscopic driving decisions that a driver undertook prior to involvement in safety-critical events. However, as the SHRP 2 NDS data consist both baseline and safety-critical events (crashes/near-crashes), it is crucial to develop volatility indices based on normal driving decisions attributable to the driving style (intentional volatility) and not the driving decisions that may have been affected due to the unsafe outcome itself (unintentional volatility), such as evasive maneuver undertaken immediately prior to a near-crash to avoid a crash event. For a detailed discussion on different components of driving volatility (intentional vs. unintentional), and the issue of reverse causality, see (Wali et al. 2018b, Wali et al. 2019). Over here, we present a brief overview of the concept in the context under discussion. Figure 2 below visualizes the speed and acceleration profiles of a baseline and crash event from the NDS data used in this study. In calculating volatility indices for baseline events, we use the entire 20-second speed and acceleration values. However, for crash and near-crash profiles (Figure 2), we employ a dynamic data censoring scheme to remove the influence of driver reactions immediately prior to a crash/near-crash from the volatility measures while retaining volatility derived from driver behavior in the seconds leading up to, but not immediately before, a crash or near-crash event. As an example, considering the speed and acceleration profiles for the sample crash event (Figure 2), 23.5 seconds (Point A in Figure 2 (right) - moment when the driver perceived the crash event and started reacting to it) of driving data are used for calculation of volatility while the rest of the data are discarded. However, note that in 14.5% of the safety-critical events (crashes/near-crashes), the driver either did not react or react after the impact. In such cases, we use driving data until the impact point (rather than using the driving data until reaction point). To fully characterize volatility in microscopic decisions, we use both acceleration and vehicular jerk-based performance measures. As deceleration profiles usually have higher variations (Kamrani et al. 2017, Arvin et al. 2019c, Arvin et al. 2019b), we use separate volatility measures for acceleration and deceleration, as well as for positive and negative vehicular jerk values, both in longitudinal and lateral dimensions. For the sake of completeness, the formulae for velocity, acceleration, and vehicular jerk are shown in Equations 1-4:

$$d = Position \tag{1}$$

$$Velocity = S = \frac{\partial d}{\partial t} \tag{2}$$

$$Acceleration = A = \frac{\partial S}{\partial t} = \frac{\partial^2 d}{\partial^2 t} \tag{3}$$

$$Vehicular\ Jerk = J = \frac{\partial A}{\partial t} = \frac{\partial^2 S}{\partial^2 t} = \frac{\partial^3 d}{\partial^3 t} \tag{4}$$

Where: $\frac{\partial}{\partial t}$ indicates derivative of a performance measure (velocity, acceleration, etc.) with respect to time, and $\partial t$ is a small change in time "$t$" (set to 0.1 seconds in this case). In this study, a total of over 9.4 million real-world driving data observations are used for calculation of volatility indices for more than 41,000 driving events. For each event, we separate acceleration and deceleration values, and calculate mean and standard deviations for each. Following (Kamrani et al. 2017, Kamrani et al. 2018, Wali et al. 2018b, Wali et al. 2019), coefficient of variation is used in the present study as a measure of volatility, i.e., the standard deviation(s) are then divided by mean values to get an estimate of relative variability in instantaneous driving decisions across different events. Finally, a similar procedure is repeated for acceleration/decelerations in lateral direction, and for vehicular jerk (both positive and negative) in longitudinal and lateral directions. As such, a total of eight different volatility measures are developed[3].

---

[3] Note that another useful measure of capturing potentially unsafe driving behaviors is "critical jerk" value (Liu and Khattak 2016, Rahman et al. 2019). In particular, speed-varying thresholds for acceleration/deceleration or vehicular jerk are defined and then instances of accelerations or vehicular jerks outside the threshold are counted as surrogate measures of safety (Bagdadi and Várhelyi 2011, Wang et al. 2015, Liu and Khattak 2016, Liu et al. 2017, Rahman et al. 2019). These studies collectively suggested that jerkiness in driving may be a useful indicator of riskier driving and higher probability of (self-reported or simulated) crash occurrence (Bagdadi and Várhelyi 2011, Rahman et al. 2019). However, the present study does not calculate the 'critical jerk' value since the focus is on capturing and quantifying the extent of variations in driving decisions. To achieve this, we use *coefficient of variation* of longitudinal and lateral accelerations and vehicular jerk as a measure of volatility in longitudinal and lateral driving decisions (Kim et al. 2016, Kamrani et al. 2017, Wali et al. 2018a, Wali et al. 2018b, Wali et al. 2019). While critical jerk has been shown as a useful measure to capture 'extreme' driving behaviors (Bagdadi and Várhelyi 2011, Wang et al. 2015, Liu and Khattak

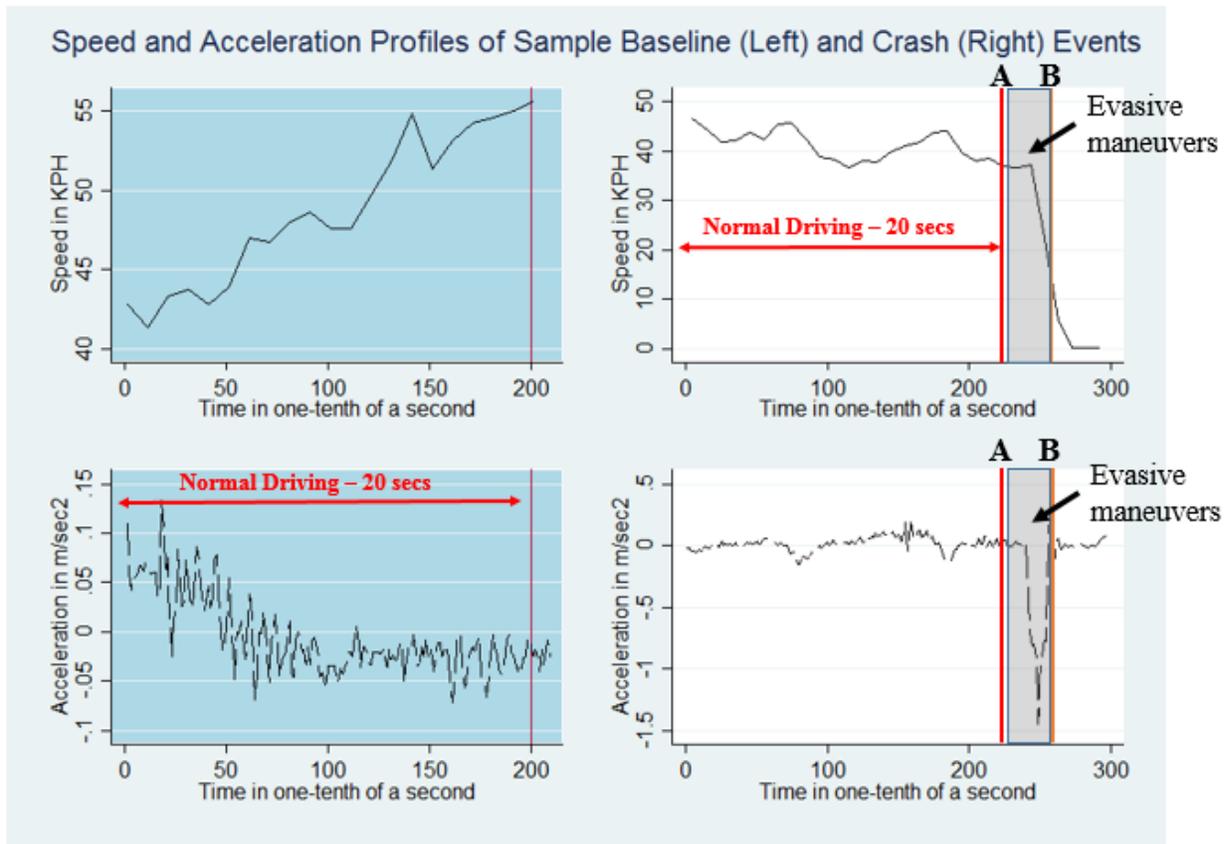

**FIGURE 2: Illustration of Dynamic Data Censoring Scheme Used for Calculation of "Intentional Volatility" in Crash/Near-Crash Events (A indicates the moment a driver starts reacting; B indicates the moment of crash impact)**

## 2.3. Statistical Models

This study examines crash propensity at school zones to understand the likelihood of crash and near-crash outcomes compared to normal driving events. Thus, the crash propensity must be considered as the likelihood of a crash or near-crash event compared to a baseline event. In particular, owing to the distinct yet related methodological concerns of scale heterogeneity and random heterogeneity, state-of-the-art discrete outcome models based on generalized mixed logit framework are developed, including multinomial logit, random parameter logit, scaled logit, hierarchical scaled logit, and a generalized hierarchical scaled logit with random parameters as special cases (see Figure 3 as explained in following text).

---

2016, Liu et al. 2017, Rahman et al. 2019), it does not explicitly quantify the extent of variations in driving decisions. That is, while it is possible that decisions of a driver in terms of vehicular jerk along a segment of a trip are within the defined thresholds or critical limits (Liu and Khattak 2016, Rahman et al. 2019), there could still be substantial variation in driving decisions relevant to safety outcomes. Using coefficient of variation allows to understand how "spread-out" the driving decisions are – it is posited that the larger the spread in acceleration or vehicular jerk values, the more safety critical it could be. As one promising avenue for future research, it would be interesting to compare coefficient of variation-based volatility measures with critical jerk or acceleration-based thresholds in predicating real-world crash or near-crash occurrence.

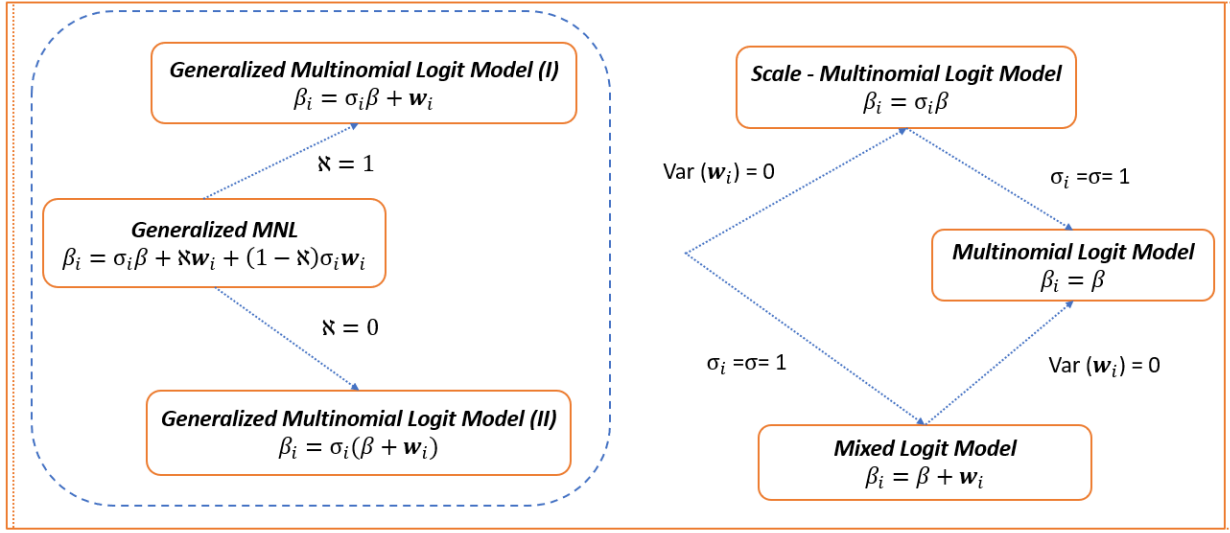

**FIGURE 3. Conceptualization of Discrete Outcome Models Considered in this Study (Framework of Generalized Multinomial Logit Model (G-MNL))**

Given the discrete nature of the event data (i.e. crash, near-crash, and baseline), a crash propensity function can be given as (Hensher and Greene 2003, Train 2003, Savolainen et al. 2011):

$$P_{ik} = \beta X_{ik} + \varepsilon_{ik} \quad (i = 1,2,3,\ldots\ldots,I; k = 1,\ldots\ldots,K) \tag{5}$$

Where $P_{ik}$ denotes crash propensity determining the probability of a crash/near-crash outcome $k$ in an event $i$, $\beta$ denotes a vector of parameters to be estimated (homogenous across observations), $X_{ik}$ denotes a vector of the explanatory variables, and $\varepsilon_{ik}$ stands for an error term assumed to have an extreme value distribution. In a standard multinomial logit setup, the probabilities of each event (i.e. crash, near-crash, and baseline) can be calculated via a simple closed-form equation (Train 2003):

$$P_n(k|X_i) = \frac{e^{(\beta X_{ik})}}{\sum_{l=1}^{K} e^{(\beta X_{il})}} \tag{6}$$

Where, $X_i$ indicates a vector of attributes related to all events ($k$ =1, 2, 3). Importantly, the MNL approach is based on two key restrictive assumptions including independent and identical distribution (IID) of $\varepsilon_{ik}$ and restrictive independence of irrelevant alternative (IIA) property. Given these key issues, the mixed logit model (i.e. random parameter MNL model) can be formulated. As of mixed logit model (RP-MNL), the crash propensity function can then be written as (Greene and Hensher 2003):

$$P_{ik} = (\beta + w_i)X_{ik} + \varepsilon_{ik} \quad (i = 1,2,3,\ldots\ldots,I; k = 1,2,3) \tag{7}$$

Where $\beta$ indicates the vector of average attribute weights of the population, $w_i$ indicates vector of the person $i$-specific deviations from the average, and the $\varepsilon_{ik}$ is assumed to follow an i.i.d. extreme value distribution. Generally, $w_i$ is assumed to have a multivariate normal distribution $(0, \sum)$, however any other appropriate distribution can also be specified for the analysis. The RP-MNL can be estimated via simulation, with the following probabilities equation feeding into the simulation protocol (Train 2003):

$$P_n(k|X_i) = \frac{1}{D}\sum_{d=1}^{D} \frac{e^{[(\beta + w^d)X_{ik}]}}{\sum_{l=1}^{K} e^{[(\beta + w^d)X_{il}]}} \tag{8}$$

The probability of each event can be calculated while taking the average of the simple logit expression over the draws D (i.e. $\{w^d\}_{d=1,2,3,\ldots,D}$ from multivariate normal distribution MVN (0, $\sum$). Note that RP-MNL (or mixed logit) captures response heterogeneity (i.e., response of likelihood of event occurrence as a function of variations in exogenous factors) through normal mixing distributions. As another promising yet more parsimonious alternative, much of the response heterogeneity could also be captured as "scale" heterogeneity – the rationale being that for some drivers the scale of idiosyncratic error term could be greater than for others. Since scale or dispersion of the error term is not identified in discrete outcome models, "scale" heterogeneity could be construed as a vector of utility weights scaled up or down proportionally for different safety events. This serves as the motivation for the scaled multinomial logit model (S-MNL) recognizing the fact that the variance/scale ($\sigma$) of the error terms in Equation 5 and 7 are normalized to one (Fiebig et al. 2010). To proceed with S-MNL, we write the standard logit model with the scale parameter however being explicit, as follows (Fiebig et al. 2010):

$$P_{ik} = \beta X_{ik} + \frac{\varepsilon_{ik}}{\sigma} \qquad (i = 1,2,3,\ldots,I; k = 1,\ldots,K = 3) \qquad (9)$$

By assuming for now that the scale parameter $\sigma$ is heterogeneous in the sample/population as $\sigma_i$ (where $i$ is an index for school zone events), multiplication of Equation 9 by $\sigma_i$ leads to the following crash propensity function:

$$P_{ik} = (\beta \sigma_i) X_{ik} + \varepsilon_{ik} \qquad (i = 1,2,3,\ldots,I; k = 1,\ldots,K) \qquad (10)$$

The scaling factor ($\sigma_i$) can consistently scale (up/down) the vector of $\beta's$ across events where the ($\sigma_i$) address a specific type of heterogeneity in the $\beta$. The S-MNL (i.e. equation 10) is more parsimonious approach compared to the RP-MNL (Equation 7). In particular, S-MNL traces scale heterogeneity whereas the RP-MNL traces random heterogeneity. Researchers have argued that much of the heterogeneity in discrete outcome models can be better captured as "scale heterogeneity" through S-MNL than as random heterogeneity as in RP-MNL. Given this contrast, we prefer to consider both S-MNL and RP-MNL for examining crash propensity at school zones. As a next variation on theme, one can also examine why the scale factor in S-MNL varies across different safety events by allowing the scale factor to vary as a function of different explanatory factors. We call this as a Hierarchical Scaled Multinomial Logit Model (HS-MNL), where the scale parameter is parameterized as:

$$\sigma_i = e^{(\sigma' + \theta z_i + \tau \varepsilon_0)} \qquad (11)$$

Where $z_i$ now indicates a vector of attributes of events, $\sigma_i$ indicates event-specific scale of idiosyncratic error which must be positive, and $\tau$ indicates standard deviation capturing the "pure scale-effect". Finally, to better examine the issue of scale heterogeneity (as in S-MNL) and random heterogeneity (as in RP-MNL) in an integrated fashion, the generalized multinomial logit (G-MNL) model can be developed which nests S-MNL and RP-MNL as special cases (Fiebig et al. 2010). The key motivation is to investigate if scale heterogeneity, normal mixing as in random parameters, or a combination of both can be a better way for describing the potential heterogeneity in crash propensity in school zones. Given the G-MNL framework, a crash propensity function can be given as:

$$P_{ik} = [\sigma_i \beta + \aleph w_i + (1 - \aleph)\sigma_i w_i]X_{ik} + \varepsilon_{ik'} \qquad (i = 1,2,3,\ldots,I; k = 1,\ldots,K) \qquad (12)$$

In the equation (12), the $\aleph$ term ranges between 0 and 1. Given equation (12) for the G-MNL, various models (i.e. MNL, RP-MNL, S-MNL, etc.) based on the values of various parameters (i.e. $\sigma_i, w_i, and \aleph$) can be derived, as shown in Figure 3. In particular, the G-MNL allows to examine scale and random heterogeneity simultaneously while disentangling the earlier from the latter. Figure 3 shows that while setting the scale parameter to one ($\sigma_i = \sigma = 1$), the G-MNL model reduces to RP-MNL, while setting $Var(w_i) = 0$ gives the scaled multinomial logit model. Whereas, the parameter ℵ, arising only in the generalized logit setup governs the proportionality between scale heterogeneity and random heterogeneity. The RP-MNL and S-MNL can be combined in a couple of ways. One way is to combine equation (7) and equation (10) - resulting in a Generalized Multinomial Logit Model I (GML I) exhibiting the form $P_{ik} = (\beta\sigma_i + w_i)X_{ik} + \varepsilon_{ik}$ (Figure 3). Alternatively, the scale parameter ($\sigma_i$) can be explicitly included in the exposition for the idiosyncratic error term in RP-MNL (equation (7)) and then multiplied by the scale parameter ($\sigma_i$) to arrive at Generalized Multinomial Logit Model II (GML II) – exhibiting the form $P_{ik} = \sigma_i(\beta + w_i)X_{ik} + \varepsilon_{ik}$. Given the above two parameterizations, the vector of utility (weights) in GML I and GML II can be written as:

$$\beta_i = \sigma_i\beta + w_i^* \qquad (13)$$

In the exposition shown in equation (13), $\sigma_i$ captures the potential heterogeneity due a pure scale-effect whereas $w_i^*$ captures random heterogeneity. The relationship between scale and random heterogeneity marks the distinction between GML I and II, where in GML II the residual heterogeneity ($w_i^*$) is proportional to scale heterogeneity ($\sigma_i$) as opposed to independence between random and scale heterogeneity in GML I (Figure 3). To estimate the event-specific scale of the idiosyncratic residual term, $\sigma_i$, a log-normal distribution with mean 1 and dispersion (standard deviation) $\tau$ is specified where the log-normal distribution is centered at 1 reflecting the fact that $\sigma_i$ must be positive (since it is a dispersion term). The broader G-MNL model transforms to GML-1 when the proportionality parameter ℵ in equation (12) approaches 1 and transforms to GML-II when ℵ approaches 0, where ℵ ∈ [0,1]. Finally, given the GMNL setup, a fully flexible Hierarchical Scaled Multinomial Logit Model with random parameters (i.e. termed as Generalized Hierarchical Mixed Logit Model – HGMNL) is considered where the scale parameter can vary across different safety events as a function of different exogenous factors (see equation 11).

## 3. DESCRIPTIVE ANALYSIS

First, the distributions of seconds of data used in calculation of driving volatility in crash and near-crash events are shown (as discussed in section 2.2). To exclude unintentional driving data immediately prior to the crash/near-crash events from calculation of driving volatility, a dynamic data censoring scheme was employed (Figure 2). Figure 4 visualizes the distributions of seconds of driving data used for calculating eight volatility indices pertaining to acceleration/deceleration and positive/negative vehicular jerk (longitudinal and lateral direction). As driver response in each of the safety-critical event is different, thus we use different pre-crash/pre-near-crash duration of data for calculation of event-specific driving volatilities. On average, around 21.51 and 22.61 seconds of driving data are used for calculating volatilities for near-crash and crash-events (Figure 4). Note that the long-left tails in the distributions shown in Figure 4 is because the duration of time-series kinematics data available for baselines and safety-critical events is not fixed (based on the data available to the authors). That is, we have few cases with less than 20 seconds of data for baselines (206 out of 32,581 baseline events). Likewise, for around 93 crash events (out of 1877 crash events) and 69 near-crash events (out of 7021 near-crash events), the duration of data available is less than 25.3 and 18.5 seconds respectively.

 Next, to examine significant differences between driving volatility in school and non-school zone locations, and between baseline and safety-critical events. Table 2 presents summary statistics of differences in mean volatilities. The differences are examined for volatility related measures based on speed, acceleration, and vehicular jerk between school zone and non-school zones. The results reveal interesting patterns about differences in average speed and speed volatility in school and non-school zone locations and among safety-critical events (baselines, near-crashes, and crashes) (Table 2). First, for all the three event types, average speeds in school zones were statistically significantly and intuitively lower than their counterparts in non-school zone locations. For instance, referring to baseline/normal driving events (Table 2), the average speed in school zones was 47.08 kph compared to the average speed of 65.78 kph in non-school zone locations (Table 2). This reveals that drivers could be more cautious in school zones in terms of their travel speeds. However, for all three event types, a statistically significant (except for crashes) and greater volatility in speeds was observed in school zones compared to non-school zones (Table 2). For example, compared to the mean speed volatility of 0.123 in non-school zones, the mean speed volatility in baseline events at school zones was greater by 40.65% (coefficient of variation of 0.173) (Table 2). The speed related findings are important in that it suggests that while drivers could be more cautious in school zones (as reflected by lower traveling speeds in school zones), the speed profiles could nonetheless exhibit substantial heterogeneity – potentially due to a relatively greater amount of complexity at school zones.

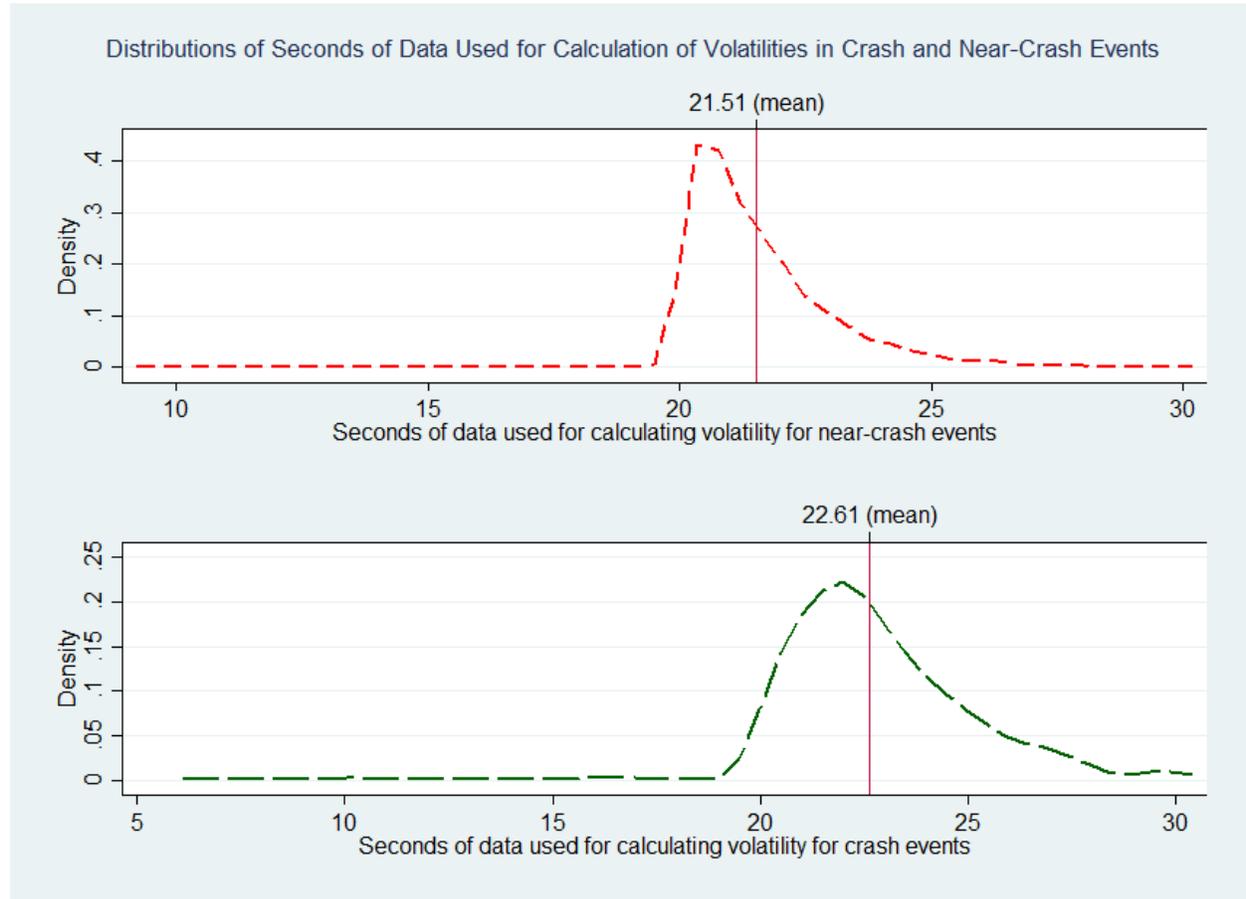

**FIGURE 4: Distribution of Seconds Used for Calculation of Intentional Driving Volatility Indices**

Coming to acceleration and vehicular jerk-based volatility measures, Table 2 shows that drivers exhibited greater intentional volatility prior to safety-critical events both at school- and non-school zones, compared to baselines. For instance, in school zones, the mean volatility in positive vehicular jerk in longitudinal direction is 1.047 in baseline events, compared to 1.157 and 1.581 for near-crash and crash events in school zones respectively (see Table 2). Next, for the four vehicular jerk-based volatility measures in longitudinal and lateral directions (Table 2), intentional driving volatility in baseline driving events was higher in school zones than at non-school zones locations (higher by a statistically significant 3.52% to 5.66% - see Table 2). Likewise, for both crash and near-crash events, intentional volatility in longitudinal positive vehicular jerk was statistically significantly greater at school zones than at non-school zone locations (Table 2). Finally, compared to non-school zone locations, drivers' volatility in longitudinal negative vehicular jerk was on-average statistically significantly greater in crashes at school zones (Table 2). Collectively, the descriptive results suggest greater volatility in school zones compared to non-school zones.

**TABLE 2: Differences in Mean Driving Volatilities at School and Non-School Zone Locations and Across Different Events**

| Event Type | Locality type | Average Speed & Speed-Based Volatility | | Acceleration/Deceleration Based Volatility Measures | | | | Vehicular Jerk Based Volatility Measures | | | |
|---|---|---|---|---|---|---|---|---|---|---|---|
| | | Average Speed (kph) | Volatility in speed | Volatility in Longitudinal Acceleration | Volatility in Longitudinal Deceleration | Volatility in Lateral Acceleration | Volatility in Lateral Deceleration | Volatility in Longitudinal Positive Vehicular Jerk | Volatility in Longitudinal Negative Vehicular Jerk | Volatility in Lateral Positive Vehicular Jerk | Volatility in Lateral Negative Vehicular Jerk |
| **Baselines** (N = 31,037) | Combined | 64.701 | 0.126 | 0.8002 | 0.7149 | 0.9783 | 0.7167 | 0.9955 | 0.826 | 0.9805 | 0.8467 |
| | Non-school zones (N = 29240) | 65.781 | 0.123 | 0.8016 | 0.7135 | 0.9747 | 0.7157 | 0.9921 | 0.8239 | 0.9785 | 0.8442 |
| | School zones (N = 1797) | 47.098 | 0.173 | 0.7731 | 0.7378 | 1.0374 | 0.7326 | 1.0473 | 0.8593 | 1.0129 | 0.8876 |
| | Mean comparison | **A** | **A** | **A** | **A** | **A** | **A** | **A** | **A** | **A** | **A** |
| | *% increase/decrease\** | *-28.40%* | *40.65%* | *-3.56%* | *3.41%* | *6.43%* | *2.36%* | *5.66%* | *4.30%* | *3.52%* | *5.14%* |
| **Near-Crashes** (N = 6705) | Combined | 54.81 | 0.348 | 0.7956 | 0.8129 | 0.9621 | 0.7636 | 1.1095 | 0.9247 | 1.1046 | 0.8971 |
| | Non-school zones (N = 6308) | 55.908 | 0.343 | 0.7955 | 0.8145 | 0.9576 | 0.7637 | 1.1065 | 0.9236 | 1.1037 | 0.8962 |
| | School zones (N = 397) | 37.507 | 0.435 | 0.7696 | 0.7875 | **1.0345** | 0.7627 | **1.1577** | 0.9421 | 1.1192 | 0.9122 |
| | Mean comparison | A | A | C | B | A | C | A | C | C | C |
| **Crashes** (N = 1770) | Combined | 32.21 | 0.63 | 0.9456 | 0.8554 | 1.0744 | 0.8959 | 1.5309 | 1.1771 | 1.3887 | 1.1899 |
| | Non-school zones (N = 1630) | 32.743 | 0.627 | 0.9415 | 0.8559 | 1.0784 | 0.8964 | 1.5265 | 1.1754 | 1.3943 | 1.187 |
| | School zones (N = 140) | 26.106 | 0.665 | 0.9939 | 0.8497 | 1.0277 | 0.8904 | **1.5814** | **1.2033** | 1.3229 | 1.224 |
| | Mean comparison | A | C | C | C | C | C | A | A | C | C |

Notes: Volatility related variables are calculated as coefficients of variation for specific kinematic performance measures, i.e., the standard deviation(s) of a performance measure (speed, acceleration, jerk) are divided by mean values to get an estimate of relative variability (see text for details). A indicates statistically significant difference at 95% confidence level between volatilities at school and non-school zone locations; B indicates statistically significant difference at 90% confidence level between volatilities at school and non-school zone locations; C indicates statistically insignificant difference at 90% confidence level between volatilities at school and non-school zone locations; (*) indicates percent increase/decrease in school-zone volatility with respect to non-school zone events; Statistics in bold indicate fields with statistically significant differences; Unit for volatility in speed is kph; Unit for volatility in jerk is $\frac{m}{sec^3}$; Unit for volatility in acceleration and deceleration is: $\frac{m}{sec^2}$; N indicates sample size of the events under a particular cluster (moving across the table).

Finally, Table 3 provides the descriptive statistics of key variables used in subsequent statistical models. As the subsequent analyses focus on school-zones, the descriptive statistics relate to school zone events and are presented for all the event types combined (baselines, near-crashes, crashes). As a result, a total of 2334 events are considered, out of which, 1797, 397, and 140 are baseline, near-crash, and crash events respectively (Table 2). Several insights can be obtained from Table 3. First, for the school-zone data at hand, both for acceleration/deceleration and vehicular jerk-based volatility measures, drivers on-average exhibited greater volatility while accelerating compared to their volatility during braking. For instance, the mean coefficient of variation for positive vehicular jerk in longitudinal direction is 1.098, whereas, the mean coefficient of variation for negative longitudinal vehicular jerk is only 0.894. Similar trend can be observed for vehicular jerk-based volatility in lateral direction, and for acceleration/deceleration-based volatility in longitudinal/lateral directions (see Table 3). Second, the distributions of volatility statistics reveal substantial variation in driving decisions across the sampled school-zone observations (baselines, near crashes, crashes) (Table 3). Regarding safety critical events, the volatility statistics show that for the sampled school-zone observations, driverss were volatile well before they anticipated the crash/near-crash outcomes (as driving data reflecting evasive maneuvers immediately prior to unsafe outcome are not used in calculating volatilities); see the volatility (coefficient of variation) statistics in Table 2.

Regarding other factors, the mean driving experience in years is around 19 years, whereas, more than 31% of the drivers had at least one driving violation in last three years. In term of exposure, as much as 70% of drivers drove less 15,000 miles a year. The data also contain information on important health related factors, with around 0.12% of drivers having angioplasty or heart bypass surgery in past. Interestingly, in around 32 events, drivers were eating without utensils, whereas in another 57 events drivers were observed using cellphone for texting. Regarding secondary tasks, the mean duration of first secondary task is 2576 milliseconds (or 2.5 seconds), and 420.1 milliseconds for the secondary task (Table 3).

**TABLE 3: Descriptive Statistics of Key Variables**

| Variable | Mean | SD | Min | Max |
|---|---|---|---|---|
| ***Key Intentional Volatility Variables*** | | | | |
| Volatility (Positive vehicular Jerk: longitudinal direction) | 1.098 | 0.438 | 0.673 | 9.163 |
| Volatility (Negative vehicular Jerk: longitudinal direction) | 0.894 | 0.289 | 0.454 | 5.978 |
| Volatility (Positive vehicular Jerk: lateral direction) | 1.050 | 0.360 | 0.559 | 6.424 |
| Volatility (Negative vehicular Jerk: lateral direction) | 0.912 | 0.310 | 0.457 | 6.016 |
| Volatility (Acceleration: longitudinal direction) | 0.794 | 0.306 | 0 | 4.594 |
| Volatility (Deceleration: longitudinal direction) | 0.753 | 0.236 | 0 | 4.711 |
| Volatility (Acceleration: lateral direction) | 1.036 | 0.354 | 0 | 4.668 |
| Volatility (Deceleration: lateral direction) | 0.747 | 0.255 | 0 | 4.113 |
| ***Driving Experience Related Factors*** | | | | |
| Driver's Education offered through private company | 0.072 | 0.259 | 0 | 1 |
| Informal driver training offered by a parent, family member or friend | 0.132 | 0.338 | 0 | 1 |
| Years of driving | 19.080 | 21.035 | 0 | 75 |
| Driver had no violation in last three years | 0.689 | 0.463 | 0 | 1 |
| Driver had one violation in last three years | 0.196 | 0.397 | 0 | 1 |
| Two or more violations in last three years | 0.114 | 0.317 | 0 | 1 |
| Annual Mileage: Less than 5000 miles | 0.145 | 0.352 | 0 | 1 |
| Annual Mileage: 5000 - 10,000 miles | 0.243 | 0.429 | 0 | 1 |
| Annual Mileage: 10,000 - 15000 miles | 0.291 | 0.454 | 0 | 1 |
| Annual Mileage: 15000 - 20,000 miles | 0.139 | 0.346 | 0 | 1 |
| Annual Mileage: 20000 - 25,000 miles | 0.069 | 0.253 | 0 | 1 |
| Annual Mileage: 25000 - 30,000 miles | 0.044 | 0.205 | 0 | 1 |
| Annual Mileage: More than 30,000 miles | 0.042 | 0.201 | 0 | 1 |
| ***Driver Health Related Factors*** | | | | |
| Driver has Astigmatism | 0.027 | 0.163 | 0 | 1 |
| Driver uses glasses for reading only | 0.099 | 0.298 | 0 | 1 |
| Driver had Bypass surgery | 0.006 | 0.074 | 0 | 1 |
| Driver had Angioplasty | 0.006 | 0.077 | 0 | 1 |
| Driver had depression | 0.044 | 0.205 | 0 | 1 |
| Driver had anxiety or panic attacks in past | 0.032 | 0.175 | 0 | 1 |
| ***Driving Behavior Related Factors*** | | | | |
| Eating without utensils | 0.014 | 0.116 | 0 | 1 |
| Driver is using cell phone/texting | 0.024 | 0.154 | 0 | 1 |
| Driver is distracted | 0.067 | 0.250 | 0 | 1 |
| ***Drivers' Secondary Task Durations*** | | | | |
| Secondary Task 1 (duration in milliseconds) | 2576.782 | 2928.319 | 0 | 15000 |
| Secondary Task 2 (duration in milliseconds) | 420.173 | 1297.670 | 0 | 14152 |
| Secondary Task 3 (duration in milliseconds) | 73.667 | 548.075 | 0 | 11211 |

Notes: SD is standard deviation; (*) For definitions of the volatility indices - see text; N = 2334 events; For definitions of different event-related variables, see InSight SHRP2 NDS website (https://insight.shrp2nds.us/data/category/events#/list).

| **TABLE 3: Descriptive Statistics of Key Variables (Continued)** | | | | |
|---|---|---|---|---|
| Variable | Mean | SD | Min | Max |
| *Legality of Maneuvers* | | | | |
| Maneuver is safe and legal | 0.883 | 0.321 | 0 | 1 |
| Maneuver is safe but illegal | 0.042 | 0.200 | 0 | 1 |
| Maneuver is unsafe and illegal | 0.056 | 0.229 | 0 | 1 |
| Maneuver is unsafe but legal | 0.020 | 0.139 | 0 | 1 |
| *Intersection-Roadway Influence* | | | | |
| Intersection influence: Uncontrolled | 0.047 | 0.212 | 0 | 1 |
| Intersection influence: Traffic Signal | 0.172 | 0.378 | 0 | 1 |
| Divided Roadway | 0.201 | 0.401 | 0 | 1 |
| Not Divided - 2 way Traffic | 0.594 | 0.491 | 0 | 1 |

Notes: SD is standard deviation; N = 2334 events; For definitions of different event-related variables, see InSight SHRP2 NDS website (https://insight.shrp2nds.us/data/category/events#/list).

## 4. MODELING RESULTS

This section presents the results of advanced statistical models detailed in section 2.3, where crash propensity in school-zones is modeled as a function of variables outlined in Table 3 (including volatility). Both acceleration/deceleration and vehicular jerk-based volatility measures were tested as explanatory factors. However, in line with previous studies (Wali et al. 2018a), preliminary analysis revealed models with vehicular jerk based volatility measures outperforming acceleration/deceleration based measures. Thus, the subsequent analysis is based on vehicular jerk-based volatility measures.

The results of statistical models are discussed next that quantify the correlations between crash propensity at school-zones and event-based volatility, after controlling for other traffic, crash, and unobserved factors. First, a series of fixed-parameter logit models are estimated in which the coefficients were held fixed across all sampled events. The fixed parameter logit models are derived from a systematic process to include most important variables (such as driving volatility related variables and others) on basis of statistical significance, specification parsimony, and intuition. Once driving volatility measures were successfully added, then other variables shown in Table 3 were included. The results of final specification of fixed-parameter logit model are shown in Table 4. As discussed earlier, presence of unobserved heterogeneity and omitted variable bias is a likely possibility, and in presence of which reliable relationships between event-based driving volatility and crash propensity cannot be established. Thus, random-parameter logit (mixed logit) models were developed where the coefficients on all the explanatory variables were tested to vary across the sampled events. Following relevant literature (Washington et al. 2010), a coefficient that either resulted in a statistically significant mean and standard deviation, or a statistically significant standard deviation was considered as a random parameter in the final specification. The results of random-parameter logit model are shown in Table 4, where a total of six explanatory variables are found to be normally distributed random parameters suggesting that the effects of these variables vary significantly across different school-zone events (see Table 4). Accounting for unobserved heterogeneity in the random parameter logit model resulted in significant improvement in model fit with a marked decrease in AIC (2208.1) and McFadden Pseudo R-square (0.583), compared to an AIC of 2267.5 and Pseudo R-square of only 0.297 for fixed-parameter counterpart (see Table 4).

**TABLE 4: Estimation Results of Fixed and Random Parameter Logit Models**

| Variable | MNL | | | | RP-MNL | | | |
|---|---|---|---|---|---|---|---|---|
| | Crash | | Near-Crash | | Crash | | Near-Crash | |
| | Beta | z-stat | Beta | z-stat | Beta | z-stat | Beta | z-stat |
| Constant | -5.094 | -10.19 | -0.728 | -1.76 | -16.413 | -6.62 | -0.794 | -1.66 |
| ***Key Intentional Volatility Variables*** | | | | | | | | |
| Volatility (Positive vehicular Jerk: longitudinal direction) | 0.937 | 3.66 | 0.367 | 1.45 | 2.530 | 3.13 | 0.063 | 0.16 |
| Volatility (Negative vehicular Jerk: longitudinal direction) | 2.055 | 5.26 | 0.727 | 1.98 | 5.888 | 5.13 | 0.991 | 2.22 |
| Volatility (Positive vehicular Jerk: lateral direction) | -0.345 | -1.05 | 1.201 | 3.61 | 0.909 | 1.07 | 2.276 | 4.3 |
| *standard deviation* | --- | --- | --- | --- | --- | --- | *0.680* | *1.95* |
| Volatility (Negative vehicular Jerk: lateral direction) | 1.617 | 4.32 | -1.264 | -3.18 | 5.561 | 4.80 | -1.975 | -3.36 |
| ***Driving Experience Related Factors*** | | | | | | | | |
| Driver's Education offered through private company | 0.846 | 2.88 | 1.049 | 1.57 | 2.358 | 2.75 | --- | --- |
| Informal driver training offered by a parent, family member or friend | --- | --- | -0.506 | -2.72 | --- | --- | -0.691 | -2.88 |
| Years of Driving | --- | --- | -0.007 | -1.94 | --- | --- | -0.010 | -2.19 |
| Driver had one violation | 0.265 | 1.11 | --- | --- | -1.208 | -0.87 | --- | --- |
| *standard deviation* | --- | --- | --- | --- | *3.760* | *2.36* | --- | --- |
| Two or more violations | 0.423 | 1.37 | 0.512 | 2.55 | 2.158 | 2.73 | 0.653 | 2.58 |
| ***Health Related Factors*** | | | | | | | | |
| Driver had Astigmatism | --- | --- | --- | --- | --- | --- | 1.387 | 1.6 |
| Driver uses glasses for reading only | 0.536 | 1.80 | --- | --- | 2.423 | 2.71 | --- | --- |
| Driver had Bypass surgery | --- | --- | 1.348 | 1.69 | --- | --- | 1.690 | 1.79 |
| Driver had depression | --- | --- | 0.365 | 1.17 | --- | --- | 0.251 | 0.64 |
| ***Driving Behavior Related Factors*** | | | | | | | | |
| Eating without utensils | 0.657 | 0.99 | --- | --- | 4.606 | 2.62 | --- | --- |
| Driver is using cell phone/texting | 1.353 | 3.12 | --- | --- | 3.913 | 2.30 | --- | --- |
| Driver is distracted | --- | --- | 3.253 | 13.23 | --- | --- | 5.415 | 7.77 |
| ***Drivers' Secondary Task Durations*** | | | | | | | | |
| Secondary Task 1 (duration in seconds) | 0.0008 | 2.49 | -0.0006 | -2.47 | 0.0001 | 1.65 | -0.0008 | -2.67 |
| Secondary Task 2 (duration in seconds) | 0.0002 | 3.20 | 0.0001 | 2.32 | 0.0006 | 2.86 | 0.0001 | 2.2 |
| ***Legality of Maneuvers*** | | | | | | | | |
| Maneuver is safe and legal | -2.248 | -7.81 | -2.533 | -12.17 | -15.112 | -6.14 | -3.080 | -7.91 |
| *standard deviation* | --- | --- | --- | --- | *10.126* | *6.23* | --- | --- |
| Maneuver is safe but illegal | -1.515 | -3.46 | -3.293 | -6.69 | --- | --- | -3.920 | -5.41 |
| *standard deviation* | --- | --- | --- | --- | *5.338* | *2.85* | --- | --- |

Notes: MNL is fixed-parameter multinomial logit; RP-MNL is random parameter multinomial logit; (---) Not Applicable; AIC is Akaike Information Criterion.

**TABLE 4: Estimation Results of Fixed and Random Parameter Logit Models (Continued)**

| Variable | MNL | | | | RP-MNL | | | |
|---|---|---|---|---|---|---|---|---|
| | Crash | | Near-Crash | | Crash | | Near-Crash | |
| | β | z-stat | β | z-stat | β | z-stat | β | z-stat |
| *Intersection-Roadway Influence* | | | | | | | | |
| Intersection influence: Uncontrolled | 1.224 | 3.16 | 2.128 | 8.66 | 1.656 | 1.43 | 2.580 | 6.56 |
| Intersection influence: Traffic Signal | --- | --- | 1.438 | 8.85 | --- | --- | 1.775 | 6.7 |
| Divided Roadway | -1.002 | -2.83 | -0.402 | -1.78 | -2.794 | -2.51 | -0.795 | -1.99 |
| *standard deviation* | --- | --- | --- | --- | *1.131* | *2.06* | --- | --- |
| Not Divided - 2 way Traffic | -0.609 | -2.66 | -0.337 | -1.87 | -2.318 | -3.19 | -0.571 | -1.97 |
| *standard deviation* | --- | --- | --- | --- | *0.895* | *1.92* | --- | --- |
| *Summary Statistics* | | | | | | | | |

| | MNL | RP-MNL |
|---|---|---|
| Log-likelihood | -1095.77 | -1060.035 |
| Mc-Fadden Pseudo $R^2$ | 0.2976 | 0.5839 |
| Number of parameters | 38 | 44 |
| N | 2319 | 2319 |
| AIC | 2267.5 | 2208.1 |

Notes: MNL is fixed-parameter multinomial logit; RP-MNL is random parameter multinomial logit; (---) Not Applicable; AIC is Akaike Information Criterion.

Next, a scaled multinomial logit model (S-MNL) is developed that tracks heterogeneity in weights of the explanatory factors by a pure scale effect (see methodology section), thus regarded as "scale heterogeneity". As opposed to a mixed logit which tracks heterogeneity by allowing coefficients to vary, the scaled multinomial logit model implies that for some safety events, the scale of idiosyncratic error term is greater, holding coefficients fixed. Interestingly, by estimating just one extra parameter (scale parameter) as in scaled logit model, the AIC significantly decreased (showing an improvement) to 2209.3 (see Table 5), compared to AIC of 2267.5 for fixed parameter logit (Table 4). Note that the scale parameter for the S-MNL is significantly greater (1.073) and statistically significant, revealing substantial heterogeneity. Importantly, the AIC statistics of random parameter logit and scaled logit are approximately equal (2208.1 vs. 2209.3) even though random parameter logit in Table 4 has five more parameter estimates than the scaled logit model (see Table 4 and 5). This important finding suggests that much of the heterogeneity in weights of explanatory factors is captured by a "pure scale effect". As a next variation on the theme, to examine why scale effect vary across events, we estimate a hierarchical scaled logit model (HS-MNL) that allows the scale effect $\sigma$ to vary across explanatory variables (second panel in Table 5). This implies that scale heterogeneity can be further decomposed into observed and unobserved portion, where the observed portion is characterized by three explanatory variables in our case: two or more violations, undivided 2-way road, and years of driving experience (see Table 5). As expected, attributing (some of) the scale heterogeneity to the observed variables (as in HS-MNL) further improved the model fit, with an AIC of 2202.4 (around 6-unit decrease in AIC compared to random parameter logit and S-MNL) (Table 4 and 5).

## TABLE 5: Estimation Results of S-MNL, HS-MNL, and H-GMNL Models

| Variable | S-MNL Crash β [z-stat] | S-MNL Near-Crash β [z-stat] | HS-MNL Crash β [z-stat] | HS-MNL Crash β [z-stat] | H-GMNL Near-Crash β [z-stat] | H-GMNL Crash β [z-stat] |
|---|---|---|---|---|---|---|
| Constant | -20.741 [-4.23] | -2.508 [-2.16] | -21.63 [-9.1] | -2.673 [-2.47] | -16.69 [-8.23] | -0.542 [-0.91] |
| **Key Intentional Volatility Variables** | | | | | | |
| Volatility (Positive vehicular Jerk: longitudinal direction) | 3.721 [3.35] | 0.887 [1.13] | 3.514 [4.65] | 1.357 [1.85] | 2.842 [3.26] | 0.127 [0.25] |
| Volatility (Negative vehicular Jerk: longitudinal direction) | 9.92 [3.920] | 2.428 [2.28] | 10.391 [7.86] | 1.811 [1.94] | 6.660 [5.40] | 1.122 [1.99] |
| Volatility (Positive vehicular Jerk: lateral direction) | -1.95 [-1.49] | 4.934 [3.63] | -2.165 [-1.73] | 4.636 [4.31] | 0.288 [0.39] | 1.765 [3.13] |
| *standard deviation* | --- | --- | --- | --- | --- | *0.800 [1.92]* |
| Volatility (Negative vehicular Jerk: lateral direction) | 6.011 [2.930] | -5.407 [-3.38] | 6.437 [4.04] | -5.014 [-3.56] | 5.123 [5.27] | -1.744 [-2.14] |
| **Proportionality parameter between scale heterogeneity & random heterogeneity** | | | | | | |
| Proportionality parameter (א) | --- | | --- | | 0.124 [1.26] | |
| **Pure Scale-Effect/Heterogeneity** | | | | | | |
| Tau ($\tau$) | 1.073 [9.91] | | 1.172 [25.29] | | 0.916 [3.65] | |
| **Observed heterogeneity in the scale factor** | | | | | | |
| Two or more violations | --- | | 0.0442 [11.42] | | -2.655 [-0.22] | |
| Not Divided - 2-way Traffic | --- | | -0.127 [-45.15] | | -0.624 [-1.60] | |
| Years of driving | --- | | 0.012 [36.54] | | -0.009 [-1.11] | |
| **Driving Experience Related Factors** | | | | | | |
| Driver's Education offered through private company | 3.154 [2.610] | --- | 3.753 [4.07] | --- | 2.0311 [2.41] | |
| Informal driver training offered by a parent, family member or friend | --- | -1.813 [-2.38] | --- | -1.98 [-2.81] | | -0.931 [-2.64] |
| Years of Driving | --- | -0.008 [-0.69] | --- | 0.008 [0.85] | | -0.011 [-1.76] |
| Driver had one violation | 2.449 [2.630] | --- | 2.788 [3.54] | --- | -2.232 [-0.96] | --- |
| *standard deviation* | --- | --- | --- | --- | *5.55 [2.04]* | --- |
| Two or more violations | 1.533 [1.220] | 1.506 [2.61] | 1.447 [1.18] | 1.499 [2.71] | 2.393 [3.50] | 0.902 [2.65] |
| **Health Related Factors** | | | | | | |
| Driver had Astigmatism | --- | 2.594 [0.85] | --- | 2.058 [0.74] | --- | 1.765 [1.30] |
| Driver uses glasses for reading only | 2.872 [2.380] | --- | 3.66 [3.80] | --- | 2.291 [2.72] | --- |
| Driver had Bypass surgery | --- | 2.405 [0.92] | --- | 1.816 [0.85] | --- | 2.112 [1.82] |
| Driver had depression | --- | 0.358 [0.38] | --- | -0.114 [-0.13] | --- | 0.396 [0.70] |

Notes: S-MNL is Scaled Multinomial Logit; HS-MNL is Hierarchical Scaled Multinomial Logit; H-GMNL is Hierarchical Scaled Multinomial Logit with Random Parameters; (---) is Not Applicable.

**TABLE 5: Estimation Results of S-MNL, HS-MNL, and H-GMNL Models (Continued)**

| Variable | S-MNL | | HS-MNL | | H-GMNL | |
|---|---|---|---|---|---|---|
| | Crash | Near-Crash | Crash | Near-Crash | Crash | Near-Crash |
| | β [z-stat] | β [z-stat] | β [z-stat] | β [z-stat] | β [z-stat] | β [z-stat] |
| *Driving Behavior Related Factors* | | | | | | |
| Eating without utensils | 1.5 [1.460] | --- | 1.724 [0.72] | --- | 3.270 [2.16] | --- |
| Driver is using cell phone/texting | 3.79 [2.12] | --- | 3.675 [2.08] | --- | 3.271 [3.30] | --- |
| Driver is distracted | --- | 9.666 [4.58] | --- | 9.715 [7.5] | --- | 6.078 [5.97] |
| *Drivers' Secondary Task Durations* | | | | | | |
| Secondary Task 1 (duration in seconds) | 0.00018 [1.460] | -0.00016 [-0.25] | 0.00018 [1.45] | -0.00003 [-0.05] | 0.00016 [1.97] | -0.0001 [-2.31] |
| Secondary Task 2 (duration in seconds) | 0.00067 [2.560] | 0.00025 [1.29] | 0.007 [3.11] | 0.0002 [1.41] | 0.0005 [2.75] | 0.0001 [1.98] |
| *Legality of Maneuvers* | | | | | | |
| Maneuver is safe and legal | -7.024 [-3.96] | -6.525 [-4.95] | -6.645 [-5.91] | -6.513 [-8.39] | -13.831 [-5.01] | -3.827 [-6.44] |
| *standard deviation* | --- | --- | --- | --- | *8.85 [5.57]* | --- |
| Maneuver is safe but illegal | -3.565 [-2.35] | -9.245 [-3.48] | -2.46 [-1.94] | -9.345 [-4.87] | -6.319 [-2.10] | -4.843 [-4.59] |
| *standard deviation* | --- | --- | --- | --- | *5.38 [2.25]* | --- |
| *Intersection-Roadway Influence* | | | | | | |
| Intersection influence: Uncontrolled | 5.415 [3.45] | 6.165 [4.71] | 4.863 [4.17] | 5.82 [7.11] | 1.770 [1.60] | 3.238 [5.21] |
| Intersection influence: Traffic Signal | --- | 4.705 [4.19] | --- | 4.99 [7.04] | | 2.323 [5.10] |
| Divided Roadway | -5.56 [-2.52] | -1.537 [-2.09] | -5.765 [-3.38] | -1.635 [-2.61] | -1.960 [2.03] | -0.878 [-1.74] |
| *standard deviation* | --- | --- | --- | --- | --- | *1.57 [2.17]* |
| Not Divided - 2-way Traffic | -2.689 [-2.66] | -1.186 [-2.15] | -3.06 [-3.76] | -1.405 [-2.84] | -1.402 [-2.22] | -0.658 [-1.75] |
| *standard deviation* | --- | --- | --- | --- | --- | *1.46 [2.56]* |
| *Summary Statistics* | | | | | | |
| Log-likelihood | -1065.66 | | -1059.19 | | -1050.07 | |
| Mc-Fadden Pseudo $R^2$ | 0.5817 | | 0.5842 | | 0.5881 | |
| Number of parameters | 39 | | 42 | | 49 | |
| N | 2319 | | 2319 | | 2319 | |
| AIC | 2209.3 | | 2202.4 | | 2198.1 | |

Notes: S-MNL is Scaled Multinomial Logit; HS-MNL is Hierarchical Scaled Multinomial Logit; H-GMNL is Hierarchical Scaled Multinomial Logit with Random Parameters; (---) is Not Applicable; AIC is Akaike Information Criterion.

Finally, to fully capture variations in effects due to "scale heterogeneity" (as in S-MNL) and random heterogeneity (as in mixed logit/random parameter logit), and the further possibility of allowing scale heterogeneity to vary across explanatory factors (as in HS-MNL), we present a fully flexible Hierarchical Scaled Multinomial Logit Model with Random Parameters (termed as Generalized Hierarchical Mixed Logit Model – H-GMNL). By simultaneously incorporating the different sources of heterogeneity, the fully flexible H-GMNL specification resulted in best-fit with the lowest AIC (the lower the better) of 2198.1 and highest Pseudo R-square of 0.5881 (see last panel in Table 5). Several important findings follow next. First, looking at the results in Table 5 for H-GMNL, the proportionality parameter between scale and random heterogeneity (א) is positive ($\rho = 0.124$), suggesting that scale and random heterogeneity for the data at hand are proportional when both are incorporated in a same model specification. As discussed in section 2.3, the א estimate of 0.124 (closer to 0) suggests that the data are closer to the G-MNL-II (Figure 3) – the variance of random heterogeneity increases with scale (see Figure 3) as opposed to random heterogeneity being invariant to the scale heterogeneity (G-MNL-I). Second, the extent of scale heterogeneity captured by $\tau$ is still around 0.916 (Table 5), suggesting significant heterogeneity due to a "pure scale-effect" even after accounting form random heterogeneity (Table 5). However, the statistical significance of observed heterogeneity in the scale-effect diminishes when random heterogeneity is simultaneously accounted for in a same model specification (see the statistical insignificance of explanatory variables tracking potential observed heterogeneity in the scale factor in Table 5). Finally, a total of six variables are found to be normally distributed random parameters in the H-GMNL, suggesting that the effects of these variables vary significantly across the sampled events in school-zones (Table 5). Finally, to better interpret the effects of key variables, marginal effects are provided for the best-fit H-GMNL model in Table 6.

## 5. DISCUSSION

### 5.1. Safety Effects of Driving Volatility

We base our discussion regarding the safety effects of driving volatility on the results obtained from H-GMNL model specification, as it resulted in the best-fit (last panel in Table 5). For the crash outcome at school zones, the parameter estimates of volatility in positive and negative vehicular jerk in longitudinal direction are both statistically significantly positive (Table 5). This implies that, compared to baseline events, larger "intentional" event-based volatility is correlated with higher likelihood of a crash event. As an example, a one-unit increase in volatility associated with positive vehicular jerk in longitudinal direction increases the probability of a crash outcome by 0.0528 units (see marginal effects in Table 6). Figure 5 visualizes the simulated event probabilities for positive vehicular jerk in longitudinal direction and shows that the probability of a crash outcome increases (direct simulated marginal effect) with an increase in magnitude of volatility (Figure 5). Importantly, the effect of a one-unit increase in volatility associated with negative vehicular jerk in longitudinal direction is almost double (marginal effect of 0.1024) the effect of positive vehicular jerk based volatility in longitudinal direction (0.0528) (Table 6). This is also reflected in the sharper curve for the effect of negative vehicular jerk in longitudinal direction on crash outcome shown in Figure 6 (compared to the curve of the effect of positive vehicular jerk-based volatility shown in Figure 5). That is, the probability of crash outcome increases significantly when volatility in negative vehicular jerk in longitudinal direction increases beyond 1.49 (see blue curve in Figure 6). Likewise, greater volatility in negative vehicular jerk in lateral direction is also associated with a marked increase in probability of observing a crash outcome (see marginal effects in Table 6 and corresponding event probabilities shown in Figure 7).

These findings are critical because it suggest that greater "intentional volatility" in positive vehicular jerk in time to crash/near-crash makes crash a more probable outcome. In addition, it shows that intentional volatility in negative vehicular jerk in longitudinal and lateral directions has more negative consequences than volatility in positive vehicular jerk in longitudinal direction. The implications are that advance warnings can be provided to the drivers in case a driver exhibits greater intentional volatility in school zones, potentially improving safety. We did not find statistically significant association between volatility in positive vehicular jerk in lateral direction and crash outcome in school zones.

Regarding the relationship of volatility with near-crash outcomes in school zones, we found statistically significant positive correlations between near-crash outcomes, and volatility measures related to negative vehicular jerk in longitudinal direction and positive vehicular jerk in lateral direction. Referring to Table 6, a one-unit increase in volatility related to positive vehicular jerk in lateral direction increases probability of near-crash outcome by 0.1287 units, compared to a statistically insignificant 0.008 units increase in near-crash likelihood with a unit increase in positive vehicular jerk in longitudinal direction. However, with a mean of 1.765 and standard deviation of 0.80, the parameter estimate for positive vehicular jerk in lateral direction is found normally distributed random parameter suggesting that the effect varies across different events (see estimates in Table 5). The above findings suggest that the effect of greater volatility in positive vehicular jerk in lateral direction can be more pronounced/more severe than its counterpart in longitudinal direction.

**TABLE 6: Marginal Effects for Best-Fit Generalized Hierarchical Mixed Logit Model**

| Variable | Generalized Hierarchical Mixed Logit Model – H-GMNL | | | | | |
|---|---|---|---|---|---|---|
| | Effects of variables in utility function of near-crash outcome | | | Effects of variables in utility function of crash outcome | | |
| | B | NC | C | B | NC | C |
| ***Key Intentional Volatility Variables*** | | | | | | |
| Volatility (Positive vehicular Jerk: longitudinal direction) | -0.0074 | **0.008** | -0.006 | -0.0388 | -0.014 | **0.0528** |
| Volatility (Negative vehicular Jerk: longitudinal direction) | -0.0536 | **0.0584** | -0.0048 | -0.074 | -0.0284 | **0.1024** |
| Volatility (Positive vehicular Jerk: lateral direction) | -0.1201 | **0.1287** | -0.0086 | -0.0037 | -0.0014 | **0.0051** |
| Volatility (Negative vehicular Jerk: lateral direction) | 0.0832 | **-0.09** | 0.007 | -0.0597 | -0.0205 | **0.0802** |
| ***Driving Experience Related Factors*** | | | | | | |
| Driver's Education offered through private company | --- | --- | --- | -0.0025 | -0.001 | **0.0035** |
| Informal driver training offered by a parent, family member or friend | 0.0077 | **-0.0084** | 0.0006 | --- | --- | --- |
| Years of Driving | 0.0104 | **-0.0109** | 0.0006 | --- | --- | --- |
| Driver had one violation | --- | --- | --- | -0.0033 | -0.0011 | **0.0044** |
| Two or more violations | -0.0071 | **0.0079** | -0.0008 | -0.0032 | -0.0021 | **0.0053** |
| ***Health Related Factors*** | | | | | | |
| Driver had Astigmatism | -0.0011 | **0.0011** | -0.0001 | --- | --- | --- |
| Driver uses glasses for reading only | --- | --- | --- | -0.0037 | -0.0009 | **0.0045** |
| Driver had Bypass surgery | -0.001 | **0.001** | 0 | --- | --- | --- |
| Driver had depression | -0.0012 | **0.0012** | -0.0001 | --- | --- | --- |
| ***Driving Behavior Related Factors*** | | | | | | |
| Eating without utensils | --- | --- | --- | -0.0011 | -0.0003 | **0.0014** |
| Driver is using cell phone/texting | --- | --- | --- | -0.0014 | -0.0007 | **0.0021** |
| Driver is distracted | -0.0247 | **0.0302** | -0.0055 | --- | --- | --- |
| ***Drivers' Secondary Task Durations*** | | | | | | |
| Secondary Task 1 (duration in seconds) | 0.0144 | **-0.0158** | 0.0014 | -0.0051 | -0.0019 | **0.007** |
| Secondary Task 2 (duration in seconds) | -0.0047 | **0.0054** | -0.0006 | -0.0031 | -0.0017 | **0.0048** |
| ***Legality of Maneuvers*** | | | | | | |
| Maneuver is safe and legal | 0.1721 | **-0.1774** | 0.0053 | 0.0035 | 0.0024 | **-0.0058** |
| Maneuver is safe but illegal | 0.0063 | **-0.0068** | 0.0006 | 0.0011 | 0.0002 | **-0.0013** |
| ***Intersection-Roadway Influence*** | | | | | | |
| Intersection influence: Uncontrolled | -0.0177 | **0.0192** | -0.0015 | -0.0009 | -0.0007 | **0.0016** |
| Intersection influence: Traffic Signal | -0.0351 | **0.037** | -0.0019 | --- | --- | --- |
| Divided Roadway | | 0.0024 | **-0.0026** | 0.0002 | 0.0031 | 0.0008 | **-0.0039** |
| Not Divided - 2 way Traffic | | 0.0002 | **-0.0008** | 0.0006 | 0.0086 | 0.0031 | **-0.0117** |

Notes: Statistics in bold indicate direct marginal effects; B is Baseline; NC is Near-Crash; C is Crash; (---) is Not Applicable

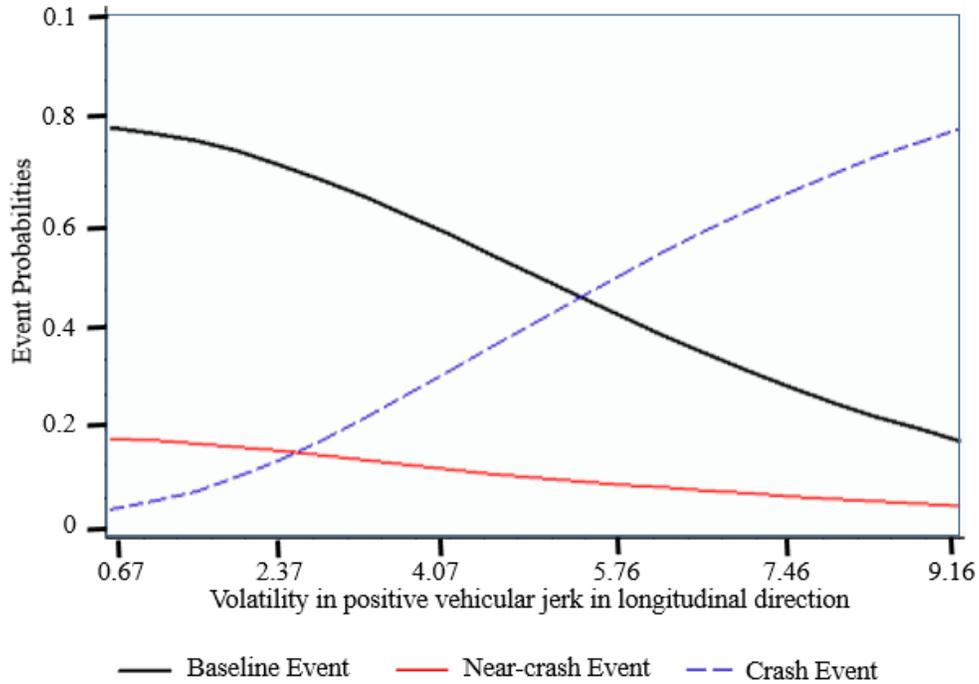

**FIGURE 5: Event Probabilities Simulated Over Events (Effect of volatility in positive vehicular jerk in longitudinal direction in crash utility function)**

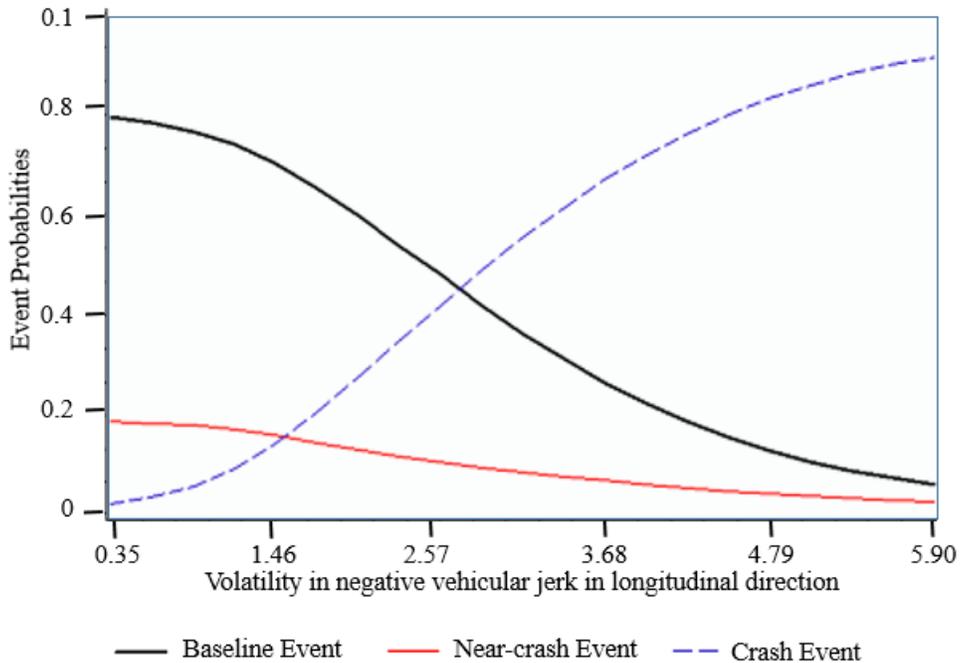

**FIGURE 6: Event Probabilities Simulated Over Events (Effect of volatility in negative**

vehicular jerk in longitudinal direction in crash utility function)

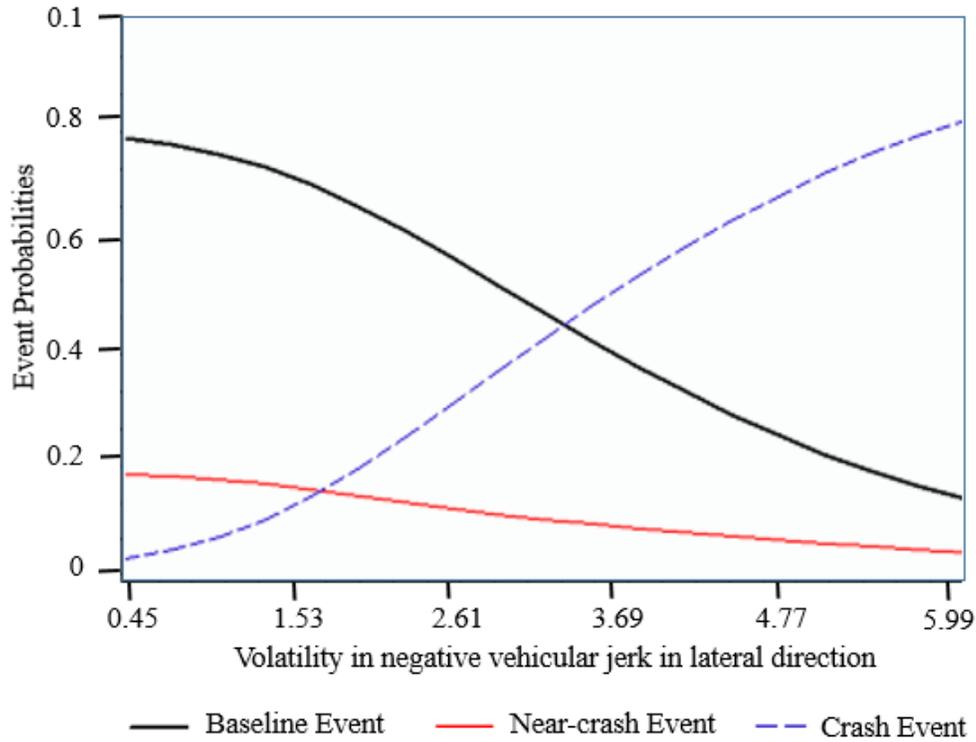

**FIGURE 7: Event Probabilities Simulated Over Events (Effect of volatility in negative vehicular jerk in lateral direction in crash utility function)**

### 5.2. Safety Effects of Other Key Factors

Due to space constraints, the safety effects of other key variables shown in Table 5 (for H-GMNL model) are not discussed but can be interpreted in a similar fashion. In particular, after accounting for scale heterogeneity, random heterogeneity, and variations in scale heterogeneity as a function of observed factors, the best-fit H-GMNL model shed light on the dependencies between crash propensity in school zones and driver experience related factors, health characteristics, driver behavior, secondary tasks and durations, legality of maneuvers, and roadway specific factors (see Table 5 and 6). For instance, if a driver had one traffic violation in past, the probability (likelihood) of getting involved into a crash increased by 0.0044 units (0.44 percentage points) (see Table 6). However, with a mean of -2.232 and relatively large standard deviation of 5.55 (Table 5), significant heterogeneity was observed in that the association was negative in 65.6% of events and positive for the rest. The substantial heterogeneity may be an outgrowth of potential "deterrence effects". However, if a driver had two or more traffic violations, their likelihood of getting involved in crash or near-crash on-average consistently increased (see the estimates for this variable in Table 5 and corresponding marginal effects in Table 6). For a detailed discussion on the relevance of deterrence theory and moral hazard in explaining the associations between previous traffic violations/convictions and crash risk, see Wali et al. (2018c) and the references therein (Wali et al. 2018c). Finally, if a driver had a bypass surgery in past, was distracted, or was engaged in secondary tasks, their likelihood of getting involved in crash/near-crash events increased (see Table 5 and 6). Other findings related to driving experience, legality of maneuvers, driving behavior, and road factors can be interpreted in a similar fashion.

# 6. SIMULATION (Predicted effects of volatility on crash propensity)

From a policy and practical standpoint, it is important to understand the potential reductions in crash frequency as a result of calmer (less volatile) driving. In this section, we present the results of simulations conducted to understand what percentages of crashes can be reduced if the volatility measures in longitudinal and lateral directions reduce. We base the simulation on the hierarchical generalized mixed logit model given its relatively best fit (Table 5). In particular, we use the best-fit hierarchical generalized mixed logit model to predict the set of outcomes for the sample followed by examining how the shares of outcomes (baselines, crashes, and near-crashes) would change if the attributes (volatility measures in this case) associated with each outcome changed. For the simulations, we specify all the three outcomes as the choice set (as opposed to a restricted choice set). To gain an in-depth understanding, we generate seven forecasting scenarios to understand the effect of reduction in four vehicular-jerk based volatility measures (in longitudinal and lateral directions) on the probability of crash as follows:

- In the first scheme, the four volatility measures (related to longitudinal and lateral directions) are decreased by 10% to 50% in increments of 10% (i.e., 10%, 20%, 30%, 40%, and 50% reduction).
- In the second scheme, the four volatility measures are also decreased by one (1) and two (2) standard deviations.

Finally, the model is simulated in each of the seven scenarios by computing the probabilities and predicting the outcome shares for the specified sample and summarizing the results – comparing them to the original base case (with no change in volatility measures). The results of the simulations are shown in Table 7 where the rows indicate the seven scenarios considered and columns indicate the percent shares and percent change in shares (and crash frequency) under different scenarios compared to baseline (no change) scenario. Regarding volatility in longitudinal direction, the simulation results reveal a potential reduction of 11 crashes (with a 10% decrease in coefficient of variation of positive vehicular jerk) to a potential reduction of 52 crashes should coefficient of variation of positive vehicular jerk decrease by 50% (Table 7). Compared to the predicted effect of volatility in positive vehicular jerk in longitudinal direction, a relatively more pronounced reduction in crash frequency is observed with a reduction in volatility in negative vehicular jerk (braking). For instance, keeping everything constant, a 10% and 50% decrease in coefficient of variation of negative vehicular jerk is associated with a reduction of 21 and 84 crashes, respectively (Table 7). Regarding volatility in lateral direction, the potential reductions in crashes ranged between 16 and 73 crashes for a 10% to 50% reduction in coefficient of variation of negative vehicular jerk (Table 7). However, the predicted effects of potential reductions in coefficient of variation of positive vehicular jerk in lateral direction seems to be indifferent (see Table 7). This finding is not surprising since this variable was statistically insignificant in the best-fit hierarchical generalized mixed logit model (see Table 5). Collectively, the above simulation findings reveal the potential of volatility measures in predicting the percentage reduction in crashes with a decrease in driving volatility.

**TABLE 7: Simulation Results for Predicted Effects of Volatility on Crash Probability (based on best-fit hierarchical generalized mixed logit model).**

| | *Predicted Effect of Volatility in Longitudinal Direction on Crash Frequency* | | | | | | | |
|---|---|---|---|---|---|---|---|---|
| | Coefficient of variation (Positive vehicular jerk) | | | | Coefficient of variation (Negative vehicular jerk) | | | |
| **Change in attribute in crash outcome function** | % Share | Number of crashes | % change in shares* | % change in crashes* | % Share | Number of crashes | % change in shares* | % change in crashes* |
| No change (baseline) | 6.414 | 149 | --- | --- | 6.436 | 149 | --- | --- |
| 10% decrease | 5.946 | 138 | -0.468% | -11 | 5.523 | 128 | -0.913 | -21 |
| 20% decrease | 5.507 | 128 | -0.928% | -21 | 4.649 | 108 | -1.787 | -41 |
| 30% decrease | 5.021 | 116 | -1.433% | -33 | 3.954 | 92 | -2.482 | -57 |
| 40% decrease | 4.609 | 107 | -1.829% | -42 | 3.372 | 78 | -3.064 | -71 |
| 50% decrease | 4.173 | 97 | -2.239% | -52 | 2.799 | 65 | -3.637 | -84 |
| 1 SD decrease | 4.751 | 110 | -1.739% | -39 | 4.081 | 95 | -2.355 | -54 |
| 2 SD decrease | 3.569 | 83 | -2.941% | -66 | 4.423 | 103 | -2.013 | -46 |
| | *Predicted Effect of Volatility in Lateral Direction on Crash Frequency* | | | | | | | |
| | Coefficient of variation (Positive vehicular jerk) | | | | Coefficient of variation (Negative vehicular jerk) | | | |
| **Change in attribute in crash outcome function** | % Share | Number of crashes | % change in shares* | % change in crashes* | % Share | Number of crashes | % change in shares* | % change in crashes* |
| No change (baseline) | 6.424 | 149 | --- | --- | 6.442 | 149 | --- | --- |
| 10% decrease | 6.399 | 148 | -0.025 | -1 | 5.716 | 133 | -0.726 | -16 |
| 20% decrease | 6.365 | 148 | -0.059 | -1 | 4.98 | 115 | -1.462 | -34 |
| 30% decrease | 6.291 | 146 | -0.133 | -3 | 4.372 | 101 | -2.07 | -48 |
| 40% decrease | 6.248 | 145 | -0.176 | -4 | 3.804 | 88 | -2.638 | -61 |
| 50% decrease | 6.151 | 143 | -0.273 | -6 | 3.296 | 76 | -3.146 | -73 |
| 1 SD decrease | 6.28 | 146 | -0.144 | -3 | 4.404 | 102 | -2.038 | -47 |
| 2 SD decrease | 6.161 | 143 | -0.263 | -6 | 3.066 | 71 | -3.376 | -78 |

Notes: (*) indicates % change in shares with respect to the baseline scenario; 1 SD is one standard deviation; 2 SD is two standard deviations; See descriptive statistics in Table 1 for the corresponding SD decrease in volatility measures considered in the simulations; The simulations are produced by changing the four volatility measures (one at a time) in the crash utility function.

## 7. CONCLUSIONS & FUTURE RESEARCH

Emerging technologies such as sensor-based monitoring, telematics, video and radar surveillance have enabled the monitoring of dynamic physical systems, generating countless terabytes of microscopic data about transport system performance. As complex layers of urban networks and digital information blanket the urban landscape, new innovative techniques to the study of major transportation challenges are needed. By harnessing the big data generated by CPS technologies, this study focused on real-world microscopic driving behavior and its relevance to school zone safety – expanding the capability, usability, and safety of dynamic physical systems through data analytics. This study focused on three key questions: (1) What pre-crash behaviors lead to risky outcomes in school zones where exposure is high, (2) What is the magnitude of driving volatility (both longitudinal and lateral) in school zones and non-school zones, and (3) How to appropriately quantify the correlations between driving volatility and crash propensity (involvement in crash and near-crash events) in school zones. To answer these questions, the study harnessed a rigorous observational study design to help compare real-world microscopic driving decisions in normal vs. unsafe outcomes at school zones. In particular, analysis of Second Strategic Highway Research Program's unique and largest Naturalistic Driving Study database of thousands of real-world driving events was conducted. In particular, more than 41,000 normal and safety-critical driving events featuring over 9.4 million real-world driving data observations are analyzed. Owing to the useful but unstructured large-scale driving data, a unique big data analytic methodology was proposed for quantifying driving volatility in microscopic real-world driving decisions. In doing so, careful attention was given to the issue of intentional vs unintentional volatility. A descriptive analysis was first conducted to spot differences between driving volatilities at school and non-school zone locations. From a methodological perspective, the study contributes by developing state-of-the-art discrete outcome models based on generalized mixed logit framework (a superset of multinomial logit, random parameter logit, scaled logit, hierarchical scaled logit, and a generalized hierarchical scaled logit with random parameters) to link driving volatility with school zone crash propensity, fully accounting for scale and random heterogeneity, with notable extension to account for the observed and unobserved components of the earlier.

In particular, among all the diverse models tested, Hierarchical Generalized Mixed Logit model resulted in best-fit, rigorously accounting for scale and random heterogeneity simultaneously. From a methodological perspective, we found significant evidence related to the presence of both scale and random heterogeneity in crash propensity at school zones. When random and scale heterogeneity are accounted for separately, discrete outcome models accounting for scale heterogeneity (as in scaled logit and hierarchical scaled logit) performed comparably to the counterparts accounting for random heterogeneity (random parameter logit) – suggesting that a more parsimonious scaled logit model can capture as much (if not more) of the heterogeneity in the data as captured by a less parsimonious (and more complicated) mixed logit model. Importantly, even after accounting for random heterogeneity, substantial heterogeneity due to a "pure scale-effect" is still observed, underscoring the importance of accounting for scale-effects. Finally, scale and random heterogeneity for the data at hand are proportional when both are present in the same model specification, as in a fully flexible H-GMNL.

The results from H-GMNL reveal that drivers exhibit greater intentional volatility prior to safety-critical events at school and non-school zones, compared to normal driving events. Likewise, for baseline events, we observed statistically significantly greater volatility at school zones compared to non-school zone locations. Regarding relationships between crash propensity and volatility at school zones, the results provide compelling evidence that an increase in volatility in positive and negative vehicular jerk in longitudinal and lateral direction increases the probability of unsafe outcomes. These findings are critical because it suggest that greater "intentional volatility" in positive vehicular jerk in time to crash/near-crash makes crash a more probable outcome. In addition, it shows that intentional volatility in negative vehicular jerk in longitudinal direction has more negative consequences than volatility in positive vehicular jerk in longitudinal direction. Given that the effects of greater volatility associated with negative vehicular jerk in longitudinal/lateral direction on crash likelihood are more pronounced and that the effect of volatility in lateral direction on safety (near-crash) outcomes is also more severe, such alerts and warnings can potentially help in improving safety.

The above volatility related findings in context of school zones have important implications for proactive safety. For instance, real-world microscopic driving decisions can be monitored in real time, and warnings and alerts can be given to drivers through emerging CPS technologies in case a driver is suspected to be more volatile in longitudinal and lateral directions (especially during braking). From a behavioral standpoint, the volatility-related findings originating from this study are based on driving data when the driver is presumably in control of the vehicle. Thus, proactive alerts and warnings can significantly enhance safety by reducing the context-specific intentional driving volatility.

There are limitations to this study pointing to subsequent future research. Foremost, speed limit management is a common strategy for school zones. If the school zones considered in this study have implemented speed limit management, that could influence driving behavior through school zones. However, information on speed limit management (and subsequently speed limits) is not available in the SHRP 2 data available to the authors. With the availability of such information in the future, incorporation of speed limit management in the analysis could provide deeper insights. Second, the present study did not analyze injury outcomes given a crash. Thus, the methodology presented in this study should be enhanced in future to analyze links between intentional driving volatility and injury outcomes in school zones. Third, direct interactions between vulnerable and motorized road users are likely in school zones. There is a need to conduct an in-depth examination of the effects of driving volatility on the potentially unsafe interactions between motorized and vulnerable road users.

## 8. ACKNOWLEDGEMENT

Sponsorship and funding for this project was provided by the Student Paper Competition: SHRP 2 Safety Data Bonanza through an award granted to B. Wali as the primary author and A. Khattak as contributory author/supervisor. The sponsorship of U.S. National Academy of Sciences in funding the first author's participation in the 2019 TRB Annual Meeting for presenting the findings of an earlier version of this paper is greatly appreciated. The assistance of Virginia Tech Transportation Institute (VTTI) data stewards' team in answering all relevant queries is acknowledged. The valuable contribution of Mr. Numan Ahmad in assisting with writing the statistical methodology (section 2.3) is also sincerely acknowledged. The authors appreciate the support provided by Ms. Whitney Atkins (VTTI) in executing the Data Use License Agreement and generously providing the breakdown of sampled events across U.S. states. Finally, the authors are thankful to two anonymous reviewers for providing several useful comments. Any opinions, findings, and conclusions or recommendations expressed in this paper are solely those of the authors.

## 9. AUTHOR CONTRIBUTIONS

The individual contributions of the authors to the paper using the relevant CRediT roles are as follows: **Behram Wali:** Conceptualization, funding acquisition, methodology, software, validation, analysis, investigation, data curation, writing – original draft, writing – review & editing, visualization, project administration. **Asad J Khattak:** Resources, supervision, writing- reviewing and editing, funding acquisition, conceptualization.